\PassOptionsToPackage{unicode}{hyperref}
\PassOptionsToPackage{hyphens}{url}
\PassOptionsToPackage{dvipsnames,svgnames,x11names}{xcolor}
\documentclass[
  12pt]{article}

\usepackage{authblk}
\usepackage{amsmath,amssymb,amsthm,mathabx}
\usepackage{iftex}
\ifPDFTeX
  \usepackage[T1]{fontenc}
  \usepackage[utf8]{inputenc}
  \usepackage{textcomp} 
\else 
  \usepackage{unicode-math}
  \defaultfontfeatures{Scale=MatchLowercase}
  \defaultfontfeatures[\rmfamily]{Ligatures=TeX,Scale=1}
\fi
\usepackage{lmodern}
\usepackage{hyperref}
\ifPDFTeX\else  
\fi
\IfFileExists{upquote.sty}{\usepackage{upquote}}{}
\IfFileExists{microtype.sty}{
  \usepackage[]{microtype}
  \UseMicrotypeSet[protrusion]{basicmath} 
}{}
\makeatletter
\@ifundefined{KOMAClassName}{
  \IfFileExists{parskip.sty}{%
    \usepackage{parskip}
  }{
    \setlength{\parindent}{0pt}
    \setlength{\parskip}{6pt plus 2pt minus 1pt}}
}{
  \KOMAoptions{parskip=half}}
\makeatother
\usepackage{xcolor}
\makeatletter
\ifx\paragraph\undefined\else
  \let\oldparagraph\paragraph
  \renewcommand{\paragraph}{
    \@ifstar
      \xxxParagraphStar
      \xxxParagraphNoStar
  }
  \newcommand{\xxxParagraphStar}[1]{\oldparagraph*{#1}\mbox{}}
  \newcommand{\xxxParagraphNoStar}[1]{\oldparagraph{#1}\mbox{}}
\fi
\ifx\subparagraph\undefined\else
  \let\oldsubparagraph\subparagraph
  \renewcommand{\subparagraph}{
    \@ifstar
      \xxxSubParagraphStar
      \xxxSubParagraphNoStar
  }
  \newcommand{\xxxSubParagraphStar}[1]{\oldsubparagraph*{#1}\mbox{}}
  \newcommand{\xxxSubParagraphNoStar}[1]{\oldsubparagraph{#1}\mbox{}}
\fi
\makeatother

\usepackage{longtable,booktabs,array}
\usepackage{calc} 
\usepackage{etoolbox}
\makeatletter
\patchcmd\longtable{\par}{\if@noskipsec\mbox{}\fi\par}{}{}
\makeatother
\IfFileExists{footnotehyper.sty}{\usepackage{footnotehyper}}{\usepackage{footnote}}
\makesavenoteenv{longtable}
\usepackage{graphicx}
\makeatletter
\def\maxwidth{\ifdim\Gin@nat@width>\linewidth\linewidth\else\Gin@nat@width\fi}
\def\maxheight{\ifdim\Gin@nat@height>\textheight\textheight\else\Gin@nat@height\fi}
\makeatother
\setkeys{Gin}{width=\maxwidth,height=\maxheight,keepaspectratio}
\makeatletter
\def\fps@figure{htbp}
\makeatother

\makeatletter
\@ifpackageloaded{caption}{}{\usepackage{caption}}
\AtBeginDocument{%
\ifdefined\contentsname
  \renewcommand*\contentsname{Table of contents}
\else
  \newcommand\contentsname{Table of contents}
\fi
\ifdefined\listfigurename
  \renewcommand*\listfigurename{List of Figures}
\else
  \newcommand\listfigurename{List of Figures}
\fi
\ifdefined\listtablename
  \renewcommand*\listtablename{List of Tables}
\else
  \newcommand\listtablename{List of Tables}
\fi
\ifdefined\figurename
  \renewcommand*\figurename{Figure}
\else
  \newcommand\figurename{Figure}
\fi
\ifdefined\tablename
  \renewcommand*\tablename{Table}
\else
  \newcommand\tablename{Table}
\fi
}
\@ifpackageloaded{float}{}{\usepackage{float}}
\floatstyle{ruled}
\@ifundefined{c@chapter}{\newfloat{codelisting}{h}{lop}}{\newfloat{codelisting}{h}{lop}[chapter]}
\floatname{codelisting}{Listing}

\makeatother
\makeatletter
\@ifpackageloaded{caption}{}{\usepackage{caption}}
\@ifpackageloaded{subcaption}{}{\usepackage{subcaption}}
\makeatother

\ifLuaTeX
  \usepackage{selnolig}  
\fi
\usepackage[]{natbib}
\usepackage{bibunits}
\renewcommand{\harvardurl}[1]{\url{#1}} 
\defaultbibliographystyle{agsm}
\defaultbibliography{refs}  
\usepackage{bookmark}

\IfFileExists{xurl.sty}{\usepackage{xurl}}{} 
\hypersetup{
  pdftitle={Title},
  pdfauthor={Author 1; Author 2},
  pdfkeywords={3 to 6 keywords, that do not appear in the title},
  colorlinks=true,
  linkcolor={blue},
  filecolor={Maroon},
  citecolor={Blue},
  urlcolor={Blue},
  pdfcreator={LaTeX via pandoc}}

\usepackage{pdflscape}
\usepackage{booktabs}  
\usepackage{multirow}

\newcommand{\var}{\operatorname{var}}

\newcommand{\cov}{\operatorname{cov}}
\newcommand{\anon}{1}
\newcommand{\vh}{\widehat{\mathrm{var}}}      
\newcommand{\ch}{\widehat{\mathrm{cov}}}
\newcommand{\bG}{\boldsymbol{\widehat{\beta}_G}}
\newcommand{\ths}{\boldsymbol{\widehat{\theta}_s}}
\newcommand{\Xt}{\boldsymbol{\widetilde{X}}}               
\newcommand{\T}{^{\top}}
\newcommand{\vecop}{\mathrm{vec}}
\newcommand{\M}{\boldsymbol{M}}
\newcommand{\Ip}{I_p}

\newtheorem{assumption}{Assumption}

\begin{document}
\begin{bibunit}

\def\spacingset#1{\renewcommand{\baselinestretch}%
{#1}\small\normalsize} \spacingset{1}


\if1\anon
{
  \title{\bf GENMETA-Lite: A Fast and Flexible Approach to Federated Learning with Disparate Covariate Information Across Studies}
  \author[1]{Emily L. Norton}
  \author[2]{Peter Kraft}
  \author[3]{Kristen D. Brantley}
  \author[2]{Thomas Ahearn}
  \author[4]{Montserrat Garcia-Closas}
  \author[5]{James M. Hodge}
  \author[6]{A. Heather Eliassen}
  \author[1]{Nilanjan Chatterjee\thanks{Corresponding author. Email: nchatte2@jhu.edu}}

  \affil[1]{Department of Biostatistics, Johns Hopkins  Bloomberg School of Public Health, Baltimore, Maryland, USA.}
  \affil[2]{Division of Cancer Epidemiology and Genetics, National Cancer Institute, Bethesda, Maryland, USA.}
  \affil[3]{Department of Medical Oncology, Dana-Farber Cancer Institute, Boston, Massachusetts, USA.}
  \affil[4]{Division of Genetics and Epidemiology, Institute of Cancer Research; Cancer Epidemiology and Prevention Research Unit, Institute of Cancer Research and Imperial College, London, UK.}
  \affil[5]{Department of Population Science, American Cancer Society, Atlanta, Georgia, USA.}
  \affil[6]{Department of Epidemiology, Harvard T.H. Chan School of Public Health, Boston, Massachusetts, USA.}
  
  \maketitle
} \fi

\if0\anon
{
  \bigskip
  \bigskip
  \bigskip
  \begin{center}
    {\LARGE\bf GENMETA-Lite: A Fast and Flexible Approach to Federated Learning with Disparate Covariate Information Across Studies}
\end{center}
  \medskip
} \fi

\bigskip
\newpage
\begin{abstract}
Meta-analysis is foundational to federated learning, where one learns parameters by combining information across studies without sharing individual-level data. Handling missing and disparate covariates across studies is a major issue due to potential introduction of bias and decreased efficiency. We previously proposed a generalized meta-analysis estimator (GENMETA) to build a unified regression model using information on parameter estimates from sub-models and a reference population to estimate the joint covariate distribution. Building on that work, we present a linear estimator (GENMETA-Lite) as a practical, non-iterative method for multivariate meta-analysis, extended to handle heterogeneity in covariate distributions across studies, coarsened covariates, and model selection. Simulations demonstrate that our estimator has robust performance compared to estimates from na\"ive inverse-variance weighted meta-analysis or multiple imputation with individual-level data ($\sim$4 to \textgreater100-fold reduction in mean squared error), even when underlying assumptions of linearity and homogeneous distributions are violated. We apply the method to fit a breast cancer risk prediction model with data from five studies (N = 313,851). Our study shows GENMETA-Lite is a robust federated learning approach for building regression models by combining information across disparate studies. An R package implementing these methods is made publicly available.
\end{abstract}

\noindent%
\vfill

\newpage
\spacingset{1.8} 

\section{Introduction}\label{sec-intro}
Meta-analysis is widely used to combine parameter estimates across multiple studies when reanalysis of pooled individual-level data is infeasible due to privacy concerns, data access challenges, or computational limitations \citep{Fleiss1993}. Because meta-analysis only requires a single transfer of summary statistics (e.g. coefficient and variance estimates) rather than individual-level data, it preserves privacy and can be implemented in a federated learning setting \citep{mcmahanCommunicationEfficientLearningDeep2017,Kairouz_et_al_2021}. Additionally, meta-analysis has been shown to be asymptotically as efficient as using pooled individual-level data in many common scenarios \citep{olkinComparisonMetaAnalysisAnalysis1998,mathewEquivalenceMetaAnalysisUsing1999,linRelativeEfficiencyUsing2010}. 

Fixed-effect or random-effect meta-analysis is used when the parameters are assumed to be constant or heterogenous across studies, respectively. One commonly used method in fixed-effect meta-analysis is the inverse variance-weighted (IVW) estimator \citep{cochranCombinationEstimatesDifferent1954,Hedges_Olkin_1985,vanhouwelingenAdvancedMethodsMetaanalysis2002}, which in the general case of vectorized parameters can be written as:
\[\boldsymbol{\hat{\beta}} = \left(\sum_{s=1}^{S}\boldsymbol{\widehat{V_s}}^{-1}\right)^{-1} \sum_{s=1}^S \boldsymbol{\widehat{V_s}}^{-1} \boldsymbol{\hat{\beta_s}}\]
where \(\boldsymbol{\hat{\beta_s}}\) is the vector of regression coefficient estimates for the \(s\)th study, and \(\boldsymbol{\widehat{V_s}}\) are the estimated variance-covariance matrices. The variance of the estimator can be further estimated as:
\[\var(\boldsymbol{\hat{\beta}}) = \left(\sum_{s=1}^S \boldsymbol{\widehat{V_s}}^{-1} \right)^{-1} \]

Fixed-effect meta-analysis assumes that the regression models from each study were adjusted using the same set of covariates \citep{andersonMetaanalyzingLogisticRegression2021,beckerSynthesisRegressionSlopes2007,nieminenApplicationStandardizedRegression2022}. However, covariates missingness at the study-level may occur by design due to different study aims or administrative limitations, or at the model-fitting stage due to inconsistent covariate adjustment \citep{debrayEvidenceSynthesisPrognosis2019}. If correlated covariates are differentially missing across study models, this will result in biased parameter estimates, and a standard fixed-effect multivariate meta-analysis would be inappropriate.

Several methods have been proposed to combine regression estimates across models fit with disparate covariates (i.e. ``reduced'' models) into a final model that contains all desired covariates (i.e. ``full'' model). One framing, known as synthesis analysis, aims to combine univariate model coefficients into a full multivariate model, for which a series of methods have been proposed (reviewed in \citet{noonSynthesisMethodsMetaAnalysis}). To combine estimates from more complex reduced models, bias correction terms have been proposed, but these require either access to individual-level data with outcome information \citep{rayMetaanalysisImbalanceMeasurement2022,yoneokaSynthesisClinicalPrediction2015} or, when only summary statistics are required, the proposed method becomes intractable for models with more than three covariates \citep{yoneokaSynthesisLinearRegression2017}. Another method,  GENMETA \citep{Kundu2019,Zhao2023}, based on generalized method of moments \citep{hansenLargeSampleProperties1982,imbensGeneralizedMethodMoments2002}, has been proposed for complex, non-linear settings. GENMETA requires only summary statistics from each study as well as a joint distribution of all risk factors estimated in a reference population. An iteratively reweighted least-squares algorithm is used to optimize a complex function that links the full and reduced model parameters. While this estimator is efficient and enjoys consistency and asymptotic normality properties, because it is optimizing over a non-convex function it often struggles to converge as the number of studies or covariates grow.

Thus, motivated by the data example discussed below, we propose a more practical estimator called ``GENMETA-Lite.'' This closed-form linear estimator approximates the GENMETA procedure while eliminating convergence issues. We show that it has only moderate bias even when the linearity assumption is violated. Furthermore, we have implemented additional functionalities to handle covariate shifts between the studies and the reference population, coarsened variables (i.e. measurement error), and model selection. Since GENMETA-Lite requires only model summary statistics and covariate covariance matrices estimated in the studies and/or a reference population, this estimator can be applied in a privacy-preserving, federated learning context. We also make available a user-friendly R package implementing the different functionalities of GENMENTA-Lite.

Our article is organized as follows. In section \ref{sec:motiv_ex} we describe the motivation for the new method, driven by the formation of a large, diverse cohort consortium aimed at developing a universal breast cancer risk prediction tool. In section \ref{sec:meth}, we introduce our new method, GENMETA-Lite. Section \ref{sec:sim} details the extensive simulation experiments used to investigate its performance relative to other off-the-shelf methods in the face of common data challenges. Section \ref{sec:RWD} provides results from a real-world data example using a subset of cohorts from the cohort consortium. Section \ref{sec-conc} summarizes our method, its performance and practicality, and proposed future work.

\section{Motivating example}\label{sec:motiv_ex}

In the U.S., breast cancer is the most common and the second deadliest cancer among women \citep{siegelCancerStatistics20242024}, but early detection via mammography can reduce its mortality \citep{marmotBenefitsHarmsBreast2013,nelsonEffectivenessBreastCancer2016}. Personalized risk prediction, based on someone's lifestyle, demographics, and personal and family history, aims to provide more accurate, individualized estimates of risk so screening strategies and lifestyle interventions may be tailored to their needs \citep{chenProjectingAbsoluteInvasive2006, gailProjectingIndividualizedProbabilities1989}.  Model-based projections have shown that personalized risk-based screening can reduce cost burden and psychological harms associated with false positives while maintaining the benefits of screening \citep{Alagoz2026,Khan2021}, and multiple randomized trials are ongoing to evaluate clinical effectiveness of risk-stratified screening \citep{delaloge135PRealtimeGenotypingbased2022,essermanRiskBasedVsAnnual2025}. 

Although many breast cancer risk prediction models have been developed (reviewed in \citealt{reSystematicReviewPrediction2025}), a model that contains a more comprehensive set of risk factors that can be universally applied to any woman, regardless of her race or ethnicity, may  improve model calibration and discrimination equitably and ultimately improve precision prevention of breast cancer. Recent evidence suggests that breast cancer risk factor associations are likely consistent across racial and ethnic groups, presenting a unique opportunity for a unified model for women from all racial and ethnic groups \citep{hursonRiskFactorsBreast2024,mccarthyValidationBreastCancer2021}. To address this gap, the Breast Cancer Risk Prediction Project (BCRPP), administered by the National Cancer Institute, was assembled (\url{https://epidataplatforms.cancer.gov/bcrpp/#home}; \citealt{brantleyPerformanceGeneralpopulationBreast2026}).

The BCRPP is a consortium of 29 cohorts, including \textgreater2.1 million female participants and \textgreater150,000 breast cancer cases (Table \ref{table:BCRPP}). Cohorts represent diverse populations from North America, Europe, and Australia. More than 100 variables related to ``core'' risk factors (i.e. demographics, anthropometrics, lifestyle, personal and family history, and reproductive history), mammographic screening, and breast cancer diagnosis have been recorded. Additionally, in the future, genetic information for many of the cohorts is expected to become available through participation in the Confluence study \citep{garcia-closasm.ConfluenceUncoveringBreast2020}, a large consortium of genome-wide association studies (GWAS) of breast cancer.  

Significant data challenges arise when working with such a large consortium. For example, because each cohort has its own study protocol, extensive covariate harmonization is required, and data are missing by design in cohorts that did not record all covariates included in the BCRPP data dictionary. Further, due to data sharing restrictions including the General Data Protection Regulation (GDPR) and the EU Data Act, not all cohorts can share individual-level data. On top of these data challenges, our model development workflow will likely require fitting many different models, e.g., some with and without interactions with race, menopause status, etc. Given the number of covariates and studies we are working with, we needed a new method to handle complex patterns of missingness that is practical, reliable, fast, and only requires summary statistics.

\section{Methods: GENMETA-Lite}\label{sec:meth}

\subsection{Set up and notation}
GENMETA-Lite was designed to be applied when a researcher desires the regression model parameters for a ``full'' model $\boldsymbol{\beta}$ ($p \times 1$ dimension) that correspond to all covariates of interest $\boldsymbol{X}$. However, the researcher only has access to estimates of regression model parameters for ``reduced'' models $\boldsymbol{\hat{\theta}_s}$ ($k_s \times 1$ dimension) which correspond to subsets of covariates $\boldsymbol{X_s}$ ($n_s \times k_s$ dimension) observed in independent studies $s \in \{1, \dots, S\}$. In order to relate the reduced model estimates $\boldsymbol{\hat{\theta}_s}$ to the full model parameters $\boldsymbol{\beta}$, the researcher will also need $\mathbb{E}[\boldsymbol{X_s}^\top \boldsymbol{X_s}]$ and $\mathbb{E}[\boldsymbol{X}^\top \boldsymbol{X}]$. These quantities may be estimated in a study, if one contains all covariates in the full model $\boldsymbol{X}$, or from a reference sample of moderate size ($n_{\text{ref}} \times p$ dimension). In the derivations below, we also refer to $\boldsymbol{y_1}, \dots, \boldsymbol{y_s}$, which are vectors of the observed outcome (each with $n_s$ x $1$ dimension) in the \(S\) studies, though the researcher does not need individual-level outcome data if the reduced model coefficients  $\boldsymbol{\hat{\theta}_s}$ are already available. Note that the reference sample does not require linkage to the outcome variable. 

\subsection{Assumptions}
\begin{assumption}\label{ass:same_distrib}
Assume that the full set of covariates \(\boldsymbol{X}\) are identically distributed for all studies and the reference population. 
\end{assumption}

\begin{assumption}\label{ass:linear_full_model} 
Assume a linear full model of the form
\(\boldsymbol{y} = \boldsymbol{X}\boldsymbol{\beta} + \boldsymbol{\epsilon}\), for \(\mathbb{E}[\epsilon] =0\).
\end{assumption}

\subsection{Derivation of the GENMETA-Lite estimator} \label{sec:obj_fn}

By Assumption \ref{ass:linear_full_model}, when \(\boldsymbol{X}^\top \boldsymbol{X}\) is invertible, an ordinary least squares solution exists for the full model with the underlying population parameter defined as $\boldsymbol{\beta} = \mathbb{E}[\boldsymbol{X}^\top \boldsymbol{X}]^{-1} \mathbb{E}[\boldsymbol{X}^\top \boldsymbol{y}]$. Similarly, for each independent (i.e. no shared participants) study (\(s \in {1,...,S}\)),  when \(\boldsymbol{X_s}^\top \boldsymbol{X_s}\) is invertible, an ordinary least squares solution exists for each reduced model such that $\boldsymbol{\hat{\theta}_s} \xrightarrow{P} \boldsymbol{\theta_s} = \mathbb{E}[\boldsymbol{X_s}^\top \boldsymbol{X_s}]^{-1} \mathbb{E}[\boldsymbol{X_s}^\top \boldsymbol{y_s}] \quad \text{as } n_s \to \infty$, where $\xrightarrow{P}$ denotes convergence in probability.

By Assumption \ref{ass:same_distrib}, we can leverage the estimates from the reduced models ($\boldsymbol{\hat{\theta}_s}$) to generate more efficient estimates for the parameters in the full model
(\(\boldsymbol{\beta}\)) by relating the two:
$\boldsymbol{\theta_s} = \mathbb{E}[(\boldsymbol{X_s}^\top \boldsymbol{X_s})]^{-1} \boldsymbol{\Delta_s} \mathbb{E}[\boldsymbol{X}^\top \boldsymbol{X}] \boldsymbol{\beta}$, where the matrix \(\boldsymbol{\Delta_s}\) (dimension $k_s$ x $p$)
 selects the covariates in \(\boldsymbol{X}\) observed in a particular study $s$ (i.e. \(\boldsymbol{X_s}\)). To simplify the following derivations, let \(\boldsymbol{A_s} = \mathbb{E}[(\boldsymbol{X_s}^\top \boldsymbol{X_s})]^{-1} \boldsymbol{\Delta_s} \mathbb{E}[\boldsymbol{X}^\top \boldsymbol{X}]\) such that $\boldsymbol{\theta_s} = \boldsymbol{A_s} \boldsymbol{\beta}$.

Thus, we define the following objective function:
\begin{eqnarray} \label{eq:obj_fn}
\mathbf{U}(\boldsymbol{\beta}) &=& [\boldsymbol{\hat{\theta}} - \boldsymbol{A} \boldsymbol{\beta}]^\top \boldsymbol{V}^{-1} [\boldsymbol{\hat{\theta}} - \boldsymbol{A} \boldsymbol{\beta}] 
\end{eqnarray} 
where \(\boldsymbol{\hat{\theta}} = [\boldsymbol{\hat{\theta}_1}, ..., \boldsymbol{\hat{\theta}_S}]\) is a \(K\) x \(1\) vector of all linear estimates from the reduced models, \(\boldsymbol{V}\) is a \(K\) x \(K\)
weighting matrix, and \(\boldsymbol{A} = [\boldsymbol{A_1}, ..., \boldsymbol{A_S}]\) is a \(K\) x \(p\) matrix.

Minimizing the objective function and setting
\(\boldsymbol{V} = \operatorname{var}(\boldsymbol{\hat{\theta}})\) to obtain the most efficient solution, we define the GENMETA-Lite estimator, expressed in summation form, as:
\begin{eqnarray} \label{eq:GENMETA-Lite_est}
\boldsymbol{\hat{\beta}_G} &=& \left(\sum_{s=1}^{S} \boldsymbol{A_s}^\top \boldsymbol{V_s}^{-1} \boldsymbol{A_s} \right)^{-1} \sum_{s=1}^S \boldsymbol{A_s}^\top \boldsymbol{V_s}^{-1} \boldsymbol{\hat{\theta}_s} 
\end{eqnarray}

To implement this in practice, we rely on plug-in estimators. Assuming, without loss of generality, that the covariates have been mean-centered, then  
\(\boldsymbol{\hat{A_s}} = \widehat{\operatorname{var}}(\boldsymbol{X_s})^{-1} \boldsymbol{\Delta_s} \widehat{\operatorname{var}}(\boldsymbol{X})\) and \(\boldsymbol{\widehat{V_s}} = \widehat{\operatorname{var}}(\boldsymbol{\hat{\theta}_s})\). We propose to use the reference population to obtain $\hat{A}_s$, and in Section~\ref{sec:cov_shift} we will describe strategies for dealing with a covariate shift problem that may cause the distribution of covariates to be different between the reference and study populations. Also notice that when the set of covariates is the same in the full model and all reduced models (i.e. \(\boldsymbol{X_s} = \boldsymbol{X}\) for all \(s\)), then \(\boldsymbol{\Delta_s} = \boldsymbol{A_s} = \boldsymbol{I}\) and
\(\boldsymbol{\hat{\theta}_s} = \boldsymbol{\hat{\beta_s}}\), and the formula simplifies to the standard IVW estimator.

\begin{figure}
\centering{
\includegraphics[width=6.4in,keepaspectratio]{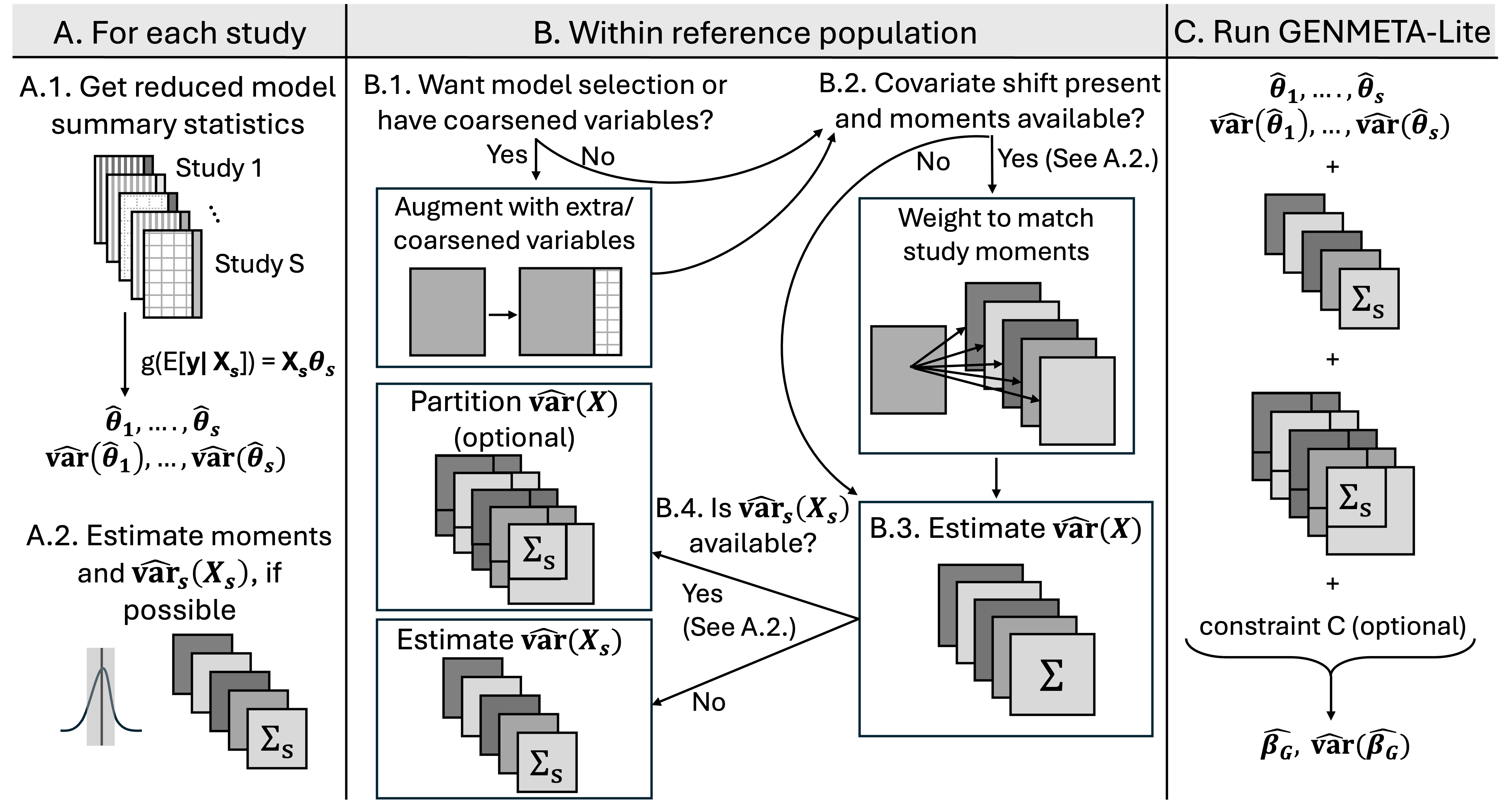}
}
\caption{\label{fig-GENMETA-Lite_workflow} Workflow for running GENMETA-Lite. First, within each study, fit reduced regression models to obtain point estimates and standard errors (A.1); if individual-level data are available, estimate the moments for each covariate and the full variance-covariance matrix across all covariates (A.2). Second, assemble a reference population that contains all covariates; if coarsened covariates are present, the reference population must be augmented with the coarsened covariates (B.1). If covariate shift is suspected and moment information is available from the studies, re-weight the reference population to match the moments of each study (B.2). Then estimate the full variance-covariance matrix (B.3). If variance-covariance matrices were estimated in each study, they can be used to partition the full variance-covariance matrix; if they are not available, the study-specific variance-covariance matrices can be estimated using the reference population (B.4). Third, use the required inputs (i.e. reduced model point estimates and standard errors, variance-covariance matrices for the study-specific covariates and all covariates in the full model), and optionally a constraint vector if model selection is desired, then run GENMETA-Lite to obtain the point estimates and standard errors for each covariate in the full model (C).}
\end{figure}

The variance of $\bG$ has two independent sources: estimation of the reduced-model
parameters $\boldsymbol{\ths}$ (from the studies) and estimation of the variance-covariance matrices that define
$\boldsymbol{A_s}$.
Writing $\boldsymbol{\M}=\big(\sum_{s=1}^S \boldsymbol{A_s}\T \boldsymbol{V_s}^{-1}\boldsymbol{A_s}\big)^{-1}$,
\begin{equation} \label{eq:GENMETA-Lite_var}
\var(\boldsymbol{\bG})=\underbrace{\boldsymbol{\M}}_{\text{if }A_s\text{ known}}
\;+\;\underbrace{\boldsymbol{\var_{\hat{A}_s}}}_{\text{from estimating }\boldsymbol{A}_s}
\end{equation}
The first term $\boldsymbol{\M}$ is the efficient meta-analysis (``sandwich'') variance obtained when
$\boldsymbol{A_s}$ are treated as fixed. The second term $\boldsymbol{\var_{\hat{A}_s}}$ accounts for the variance associated with estimating the
variance-covariance matrix the define $\boldsymbol{A_s}$ in a finite sample. Details of the derivation of the variance terms and their plug-in estimates are provided in the Supplementary Materials (Section~\ref{sec:var_estimator_deriv}). Figure \ref{fig-GENMETA-Lite_workflow} summarizes the user inputs and workflow to calculate the GENMETA-Lite estimates, along with a few optional functionalities described below.

\subsection{Additional functionalities for the GENMETA-Lite estimator}

\subsubsection{Mitigating covariate shift}  \label{sec:cov_shift}  
The procedure described above requires estimation of matrices of the form $\boldsymbol{\hat{A_s}} = \widehat{\operatorname{var}}(\boldsymbol{X_s})^{-1} \boldsymbol{\Delta_s} \widehat{\operatorname{var}}(\boldsymbol{X})$. Ideally, components of $\boldsymbol{A_s}$ should be estimated using a reference data from the same population underlying the corresponding study $s$. However, such study-specific reference data with all required covariate data are typically not available. We propose using a central reference dataset, such as one derived from national health survey data, to estimate $\boldsymbol{A_s}$. Below, we propose additional techniques to account for potential differences in covariate distributions across the populations underlying the central reference data and the studies. Handling distributional shift, however, requires additional summary statistics from the participating studies.

For the estimation of each $\boldsymbol{A_s}$, we propose re-weighting the reference data so that the moments of observed covariates $\boldsymbol{X_s}$, including covariances, match with those observed for the $s$th study. In our R package, we use entropy balancing (\citealt{Hainmueller2012}; ebal R package, version 0.1-8) to estimate weights on the reference population that exactly satisfy a set of user-specified moment-balance constraints (e.g., matching the first two moments of \(age\) and the covariance between \(age\) and \(sex\) with a given study). Following \citet{Kullback1959}, these weights minimize an entropy-based distance from a set of base weights (base weights $=1/n_{\text{ref}}$ if not otherwise specified), subject to the estimated weights summing to unity. The underlying loss function is convex and is solved via Newton-Raphson (full details in Supplementary Materials Section~\ref{sec:entropy_balancing}, following \citealt{Hainmueller2012}). We opted for entropy balancing because it guarantees exact covariate balance and thus does not require balance checking. Alternative methods, such as inverse probability of sampling \citep{Horovitz_Thompson1952,Robins1994} or raking \citep{Kruithof1937}, could be used to generate weights though they do not ensure balance.

When a study provides the full variance-covariance matrix for the observed covariates, we propose to utilize that information in the estimation of $\boldsymbol{A_s}$ by partitioning the matrix $\widehat{\var}_s(\boldsymbol{X})$ as

$$
\widehat{\var}_s(\boldsymbol{X}) = \begin{bmatrix} \widehat{\var}_s(\boldsymbol{X_s}) & \widehat{\cov}(\boldsymbol{X_s},\boldsymbol{\Xt}) \\
\widehat{\cov}(\boldsymbol{X_s},\boldsymbol{\Xt})^\top & \widehat{\var}(\boldsymbol{\Xt})
\end{bmatrix}
$$

where $\widehat{\var}_s(\boldsymbol{X_s})$ is the covariance estimated in the $s$th study for all covariates available in the study, $\widehat{\var}(\boldsymbol{\Xt})$ is the variance estimated in the reweighted reference population (see details in Section \ref{sec:cov_shift} for recommendations on weighting) for all variables that are included in the full model but not available in the study, and $\widehat{\cov}(\boldsymbol{X_s},\boldsymbol{\Xt})$ is the cross-covariance estimated in the reweighted reference population for the variables that are and are not available in the study, respectively. 

When the partitioned variance-covariance matrix is used, $\boldsymbol{A_s}=\big[\,\boldsymbol{I_{k_s}}\ \ \vh_s(\boldsymbol{X_s})^{-1}\ch(\boldsymbol{X_s},\boldsymbol{\Xt})\,\big]$,
so the reference population contributes only through the cross-covariance $\ch(\boldsymbol{X_s},\boldsymbol{\Xt})$,
while the study contributes through its own $\vh_s(\boldsymbol{X_s})$. Under such estimation, the variance of the final GENMETA-Lite estimator $\boldsymbol{\hat{\beta}_G}$ requires some further adjustments (see Section~\ref{sec:var_estimator_deriv} of the Supplementary Materials).

\subsubsection{Linear constraint to shrink regression coefficient(s) to zero} \label{sec:linear_constraint}
In some settings it may be desirable to develop a full model with specific coefficients constrained to be zero. In backward model selection, for example, one may consider fitting a full model omitting a covariate $\boldsymbol{X_j}$ while preserving the other covariates. We observe that depending on the correlation structure among covariates, $\boldsymbol{X_j}$ may still have contributed to a reduced model, and hence omitting $\boldsymbol{X_j}$ from the final model does not imply eliminating it from the entire analysis.

Another use for this functionality arises when a covariate is measured precisely in some studies ($\boldsymbol{X}$) but only in coarsened, truncated, or otherwise error-prone form ($\boldsymbol{X^*}$) in others. Assuming the measurement error is non-differential, i.e., $P(\boldsymbol{Y}|\boldsymbol{X_1}, \dots, \boldsymbol{X}, \boldsymbol{X^*}) = P(\boldsymbol{Y}|\boldsymbol{X_1}, \dots, \boldsymbol{X})$ \citep{CarrollMeasurementError}, we constrain the error-prone variable's coefficient $\beta^*$ to $0$ in the full model, which excludes it from the final model while still letting it inform the adjustment used in the reduced models (demonstrated with a coarsened \textit{age} variable in Section~\ref{sec:sim}). 

We implement this functionality by applying a Lagrangian penalty to impose a linear constraint on a specific $\beta_j$ corresponding to the covariate $\boldsymbol{X_j}$ we would like to exclude from the final model. Specifically, we minimize the objective function \eqref{eq:obj_fn} subject to \(\boldsymbol{C}\boldsymbol{\beta} = \boldsymbol{0}\), for some matrix \(\boldsymbol{C}\) with dimension \(q\) x \(p\), where
\(q\) is the number of parameters being constrained (\(q<p\)). The Lagrangian formulation then becomes:
\[
\mathcal{L}(\boldsymbol{\beta}, \lambda) = (\hat{\boldsymbol{\theta}} - \boldsymbol{A}\boldsymbol{\beta} )^T\boldsymbol{V}^{-1}(\hat{\boldsymbol{\theta}} - \boldsymbol{A}\boldsymbol{\beta}) + 2\boldsymbol{\lambda}^\top \boldsymbol{C} \boldsymbol{\beta}
\]
for some penalty parameter $\boldsymbol{\lambda}$. By minimizing this function, we obtain the following constrained estimator and its variance:
\begin{eqnarray*}
\boldsymbol{\bG^{c}}=\boldsymbol{\bG}-\boldsymbol{\M C}\T\big(\boldsymbol{C\M C}\T\big)^{-1}\boldsymbol{C}\,\boldsymbol{\bG}=\boldsymbol{P}\,\boldsymbol{\bG},\qquad
\boldsymbol{P}:=\boldsymbol{I_p}-\boldsymbol{\M C}\T\big(\boldsymbol{C\M C}\T\big)^{-1}\boldsymbol{C} 
\end{eqnarray*}
\begin{eqnarray*}
\var\!\big(\boldsymbol{\bG^{c}}\big)=\boldsymbol{P}\,\var(\boldsymbol{\bG})\,\boldsymbol{P}\T
=\Big[\boldsymbol{\M}-\boldsymbol{\M C}\T\big(\boldsymbol{C\M C}\T\big)^{-1}\boldsymbol{C\M}\Big]
\;+\;\boldsymbol{P}\big(\boldsymbol{\var_{\mathrm{ref}}}+\boldsymbol{\var_{\mathrm{study}}}\big)\boldsymbol{P}\T 
\end{eqnarray*}
where $\boldsymbol{\widehat{\beta_G}}$ is the original GENMETA-Lite estimator without a constraint \eqref{eq:GENMETA-Lite_est}. Note that the constrained covariates must still be included in the reference population since it is necessary to estimate their covariance with other covariates retained in the final model. 

\subsection{Other practical issues}
A related but distinct scenario occurs when study-specific variables (e.g. assessment center) are used to fit the reduced models. In this case, there is no meaningful analog for this variable in the reference population, so its covariance cannot be estimated externally. If a covariance matrix is provided by the study, though, then some adjustments may be made. See further discussion in the Supplementary Materials (Section~\ref{sec:study-spec-vars}).

\section{Simulation Studies} \label{sec:sim}
\begin{figure}
\centering{
\includegraphics[width=6.5in,keepaspectratio]{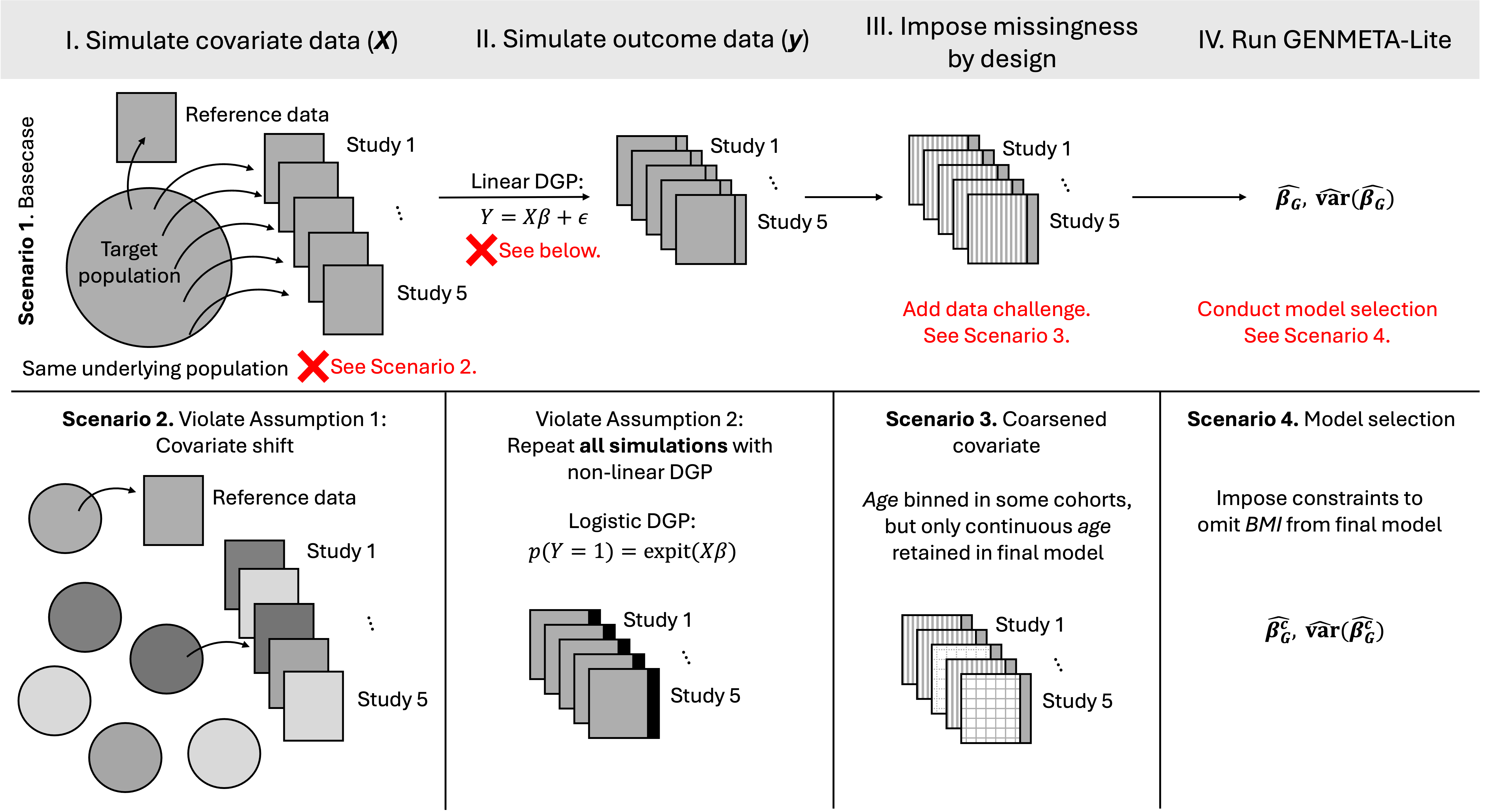}
}
\caption{\label{fig-simulations} A summary of the simulation studies conducted to evaluate the performance of the GENMETA-Lite estimator under realistic data challenges. Covariate data were simulated for five studies with samples sizes of 8,000 to 50,000, and for a reference population with sample size 5,000 (though sensitivity analyses were run for sample sizes varied from 1,000 to 100,000; see Table \ref{table:sup_sim_missingness}). Within each study, simulated covariate data were used to simulate outcomes, then structural missingness was imposed before GENMETA-Lite was run. Scenario 1 (Basecase) was used to evaluate performance when both assumptions were met and no additional data challenges were present. A red "X" and text indicate where modifications were made to simulate common data challenges: we  simulated scenarios where the assumption of the same underyling distribution was violated (Scenario 2), a coarsened covariate was present (Scenario 3), and model selection was conducted (Scenario 4). Additionally, all simulations were repeated with a logistic data generating process (DGP), violating the linearity assumption.} 
\end{figure}

We devised a set of simulation scenarios to evaluate the performance of the GENMETA-Lite estimator under common data challenges (Fig. \ref{fig-simulations}).  While realistic values were chosen for regression parameters (Table \ref{table:sup_sim_coeffs}), covariance across covariates (Fig.~\ref{fig:sup_sim_ggpairs}), and sample sizes for each study and the reference population (Table \ref{table:sup_sim_missingness}), sensitivity analyses were run to explore the effects of varying the size of the reference population ($n_{\text{ref}}=$ 1,000 to 100,000) and the magnitude of the regression coefficients (Table~\ref{table:sup_sim_coeffs}) on the bias and variance of the GENMETA-Lite estimator. Simulations were run with 100 replicates. 

\subsection{Data generating process} \label{sec:dgp}
For each of five studies and a reference population, ten covariates (five continuous, five discrete) were jointly simulated with a complex covariance structure using the SimCorrMix R package (version 0.1.1; https://github.com/AFialkowski/SimCorrMix) to produce a realistic design matrix (Table \ref{table:sup_sim_data}; Figure \ref{fig:sup_sim_ggpairs}); we have named these covariates for easier reference. Except for the case of model selection where the ``omitted'' variable was excluded, all covariates were used to simulate the outcome variable using either a linear or logistic data generating process (DGP). Structural missingness was imposed by removing all observations for a subset of covariates, which differed across studies, to mimic realistic scenarios where different variables are ``missing by design'' in each study (Table \ref{table:sup_sim_missingness}). 

\subsection{Scenarios}
For the basecase (Scenario 1), the underlying distribution of the covariates was the same for all studies and the reference population. For Scenario 2, we assumed that the study and reference population were drawn from different underlying populations, violating Assumption \ref{ass:same_distrib}. Specifically, we varied the mean and variance for \textit{age}, \textit{sex}, and \textit{smoking status}, and the correlations between \textit{BMI} and \textit{early BMI}, and between \textit{parity} and \textit{smoking status}, for each study and the reference population (Table \ref{table:sup_sim_diff_distrib}). For Scenario 3, a single variable (\textit{age}) was coarsened into 5-year bins in Studies 2 and 3, but the continuous \textit{age} variable was included in the other studies and in the final ``full'' model. For Scenario 4, model selection was simulated by omitting a single variable (\textit{BMI}) from the DGP and the final ``full'' model, though it had been included in three of the five reduced models (see missingness patterns in Table \ref{table:sup_sim_missingness}). Because we anticipate using this method in both linear and non-linear regression settings, all scenarios were run using both a linear and logistic DGP (i.e. violating Assumption \ref{ass:linear_full_model}) to simulate the outcome variable. 

\subsection{Running GENMETA-Lite}
Reduced models were fit within each study according to the DGP used to simulate the data (i.e. linear or logistic) to obtain point estimates and robust sandwich variance estimates (sandwich R package, v.3.1-1; \citealt{Huber1967,White1980,Zeileis_Graham2020}) for each covariate in the reduced models.

Depending on the simulation scenario, $\widehat{\var}(\boldsymbol{X_s})$ and $\widehat{\var}(\boldsymbol{X})$ were obtained differently. For Scenarios 1 (basecase), 3 (coarsened covariate), and 4 (model selection), since the underlying distributions for the reference population and all studies were the same, weighting was not implemented but $\widehat{\var}_s(\boldsymbol{X_s})$ was estimated in the study and was also used to build a partitioned variance matrix for $\widehat{\var}_s(\boldsymbol{X})$; for Scenarios 3 and 4, the reference population was augmented with the variable that would be omitted from the final model (i.e., Scenario 3: \textit{age} binned into 5-year intervals; Scenario 4: \textit{BMI}) prior to estimating covariance. For Scenario 2 (covariate shift), three different methods were applied to evaluate our ability to overcome the violation of Assumption \ref{ass:same_distrib}. First, in the simplest case, an unweighted reference population was used to estimate both $\widehat{\var}(\boldsymbol{X_s})$ and $\widehat{\var}(\boldsymbol{X})$. Second, assuming that each study could provide moment information, weights were estimated to balance the first two moments for the continuous variables \textit{age}, \textit{BMI}, and \textit{early BMI}, and the first moment for the categorical variables \textit{sex} and \textit{smoking status}, between the reference population and each study; covariances were also balanced. Then the weighted reference population was used to estimate both $\widehat{\var}(\boldsymbol{X_s})$ and $\widehat{\var}(\boldsymbol{X})$. Third, in the event that each study was able to share a covariance matrix in addition to moment information, we then estimated $\widehat{\var}_s(\boldsymbol{X_s})$ in each study, which was  used to build a study-specific partitioned variance matrix  $\widehat{\var}_s(\boldsymbol{X})$ by combining this with the weighted reference population described above. 

Finally, when GENMETA-Lite was run for Scenarios 3 and 4, a linear constraint was imposed which forced the coefficients for the omitted covariates (Scenario 3: \textit{age} binned into 5-year intervals; Scenario 4: \textit{BMI}) to be zero in the ``full'' model.

\subsection{Estimator performance and comparative methods}
Percent bias, empirical standard deviation of percent bias, and mean squared error (MSE) were calculated for each covariate estimate across all simulations. Additionally, to summarize performance across covariates, mean absolute percent bias and mean standard deviation of percent bias were calculated by averaging across all covariates. We compared performance of the GENMETA-Lite estimator to three off-the-shelf estimators. First, the ``gold standard'' estimator was an IVW estimator that combined full models fit in each study prior to missingness being imposing; although it is not feasible in practice, it ensures that our simulations are error-free. Second, the ``na\"ive inverse-variance weighted'' estimator ignored differential covariate adjustment by using an IVW method to combine all reduced model regression estimates that were available for a particular covariate. Third, the ``imputation'' estimator used parametric or non-parametric imputation to fill in missing data prior to fitting models and combining their estimates using an IVW method.  For more details see Supplementary Materials Section~\ref{sec:comp_estimators}. 

\subsection{Results}

\subsubsection{Scenario 1: Basecase}
\begin{figure}[!htb]
    \centering
    \includegraphics[height=0.6\textwidth]{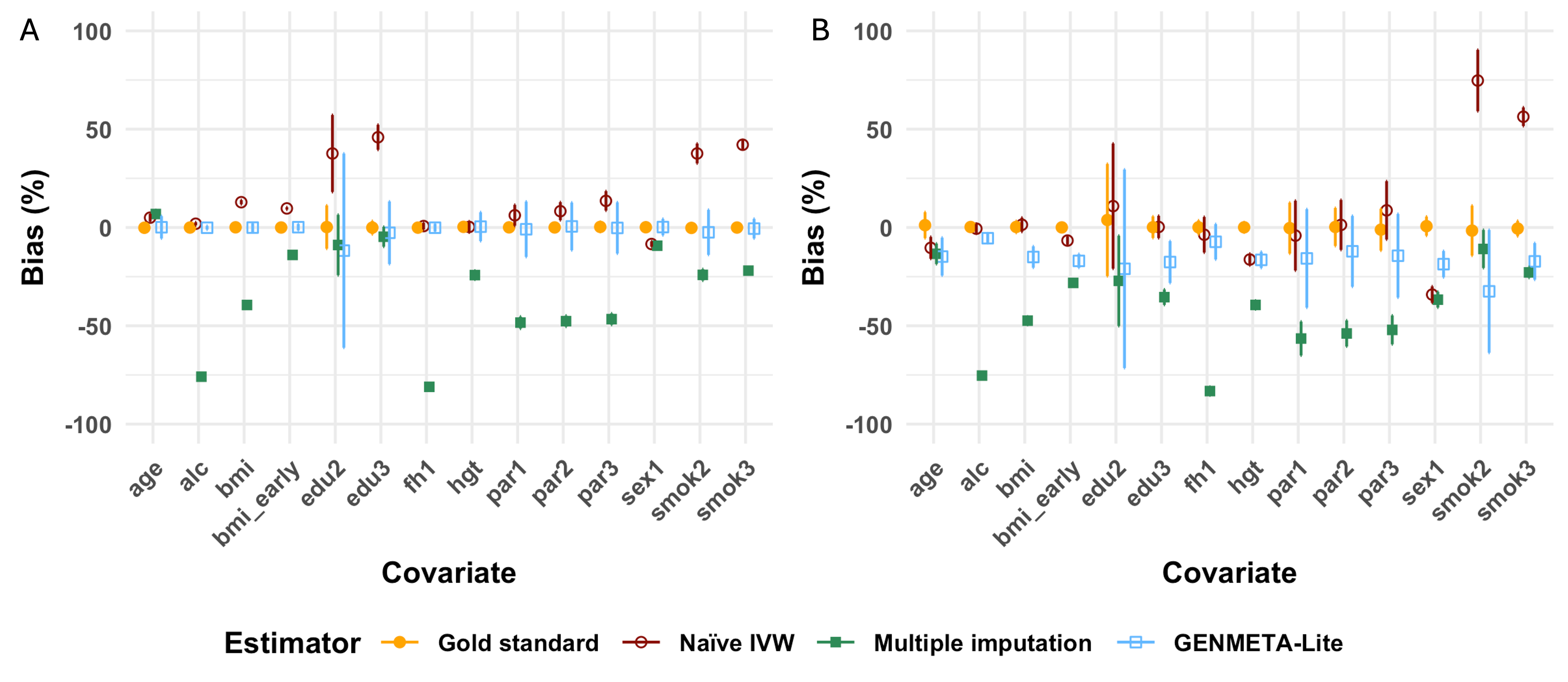}
    \caption{Mean percent bias for Scenario 1 (basecase) under (A) a linear data generating process (DGP) and (B) a logistic DGP, for the gold standard estimator (filled gold circles), the na\"ive inverse-variance weighted (IVW) meta-analysis estimator (open red circles), multiple imputation with parametric models (M=5; filled green squares), and GENMETA-Lite (open blue squares); error bars indicate empirical standard deviation of percent bias across simulations. Larger regression coefficients were used to simulate the outcome variables (see $\boldsymbol{\beta_2}$ and $\boldsymbol{\textit{OR}_2}$ in Table \ref{table:sup_sim_coeffs}). A partitioned covariance matrix, which combined covariance estimates from each study and covariance estimates from an unweighted reference population ($n_{\text{ref}} = 5000$), was used to estimate covariance across all covariates.}
    \label{fig:basecase_logistic_parametric_partitioned_var}
\end{figure}

In the basecase (Scenario 1), the GENMETA-Lite estimator controlled bias and variance well under the linear DGP (mean absolute bias: 1.5\%; mean empirical standard deviation (SD) of percent bias: 10.4\%) but showed a modest bias toward the null under the logistic DGP (mean bias: -16.1\%), with categorical variables in sparse levels (e.g. \textit{edu2}, \textit{smok2}) exhibiting the largest bias and variance under both DGPs (Fig.~\ref{fig:basecase_logistic_parametric_partitioned_var}; full results in Table \ref{table:supp_scenario_results}). Relative to GENMETA-Lite, the na\"ive IVW estimator, which ignored missing covariates, had substantially higher bias under both DGPs (mean absolute bias: 16.4\% for linear and logistic) but lower variance (mean SD: 4.2\% linear, 9.5\% logistic), while parametric and non-parametric imputation-based estimators had the largest bias of all methods evaluated (mean absolute bias: 32.4\% linear, 41.6\% logistic; Fig.~\ref{fig:basecase_logistic_pmm_partitioned_var}) despite low variance. As expected, the gold standard estimator, fit on complete data prior to imposing missingness, had negligible bias and well-controlled variance under both DGPs.

Sensitivity analyses (Table~\ref{table:supp_scenario_results}; Figs.~\ref{fig:sens_analy_bias_sd}--\ref{fig:sens_analy_coverage_rates}) showed that GENMETA-Lite's bias was insensitive to the size of the reference population, while its standard error decreased asymptotically as the reference population grew; 95\% confidence interval (CI) coverage was close to nominal for the linear DGP but below nominal for the logistic DGP. Using an unweighted reference population instead of weighted, partitioned covariance matrix did not meaningfully change bias or variance when the reference and study populations shared the same underlying distribution. Because this assumption cannot be verified in practice, we recommend the weighting and partitioning approach as a low-risk default.

\subsubsection{Scenario 2: Covariate shift}
\begin{figure}[!htb]
    \centering
    \includegraphics[height=0.6\textwidth]{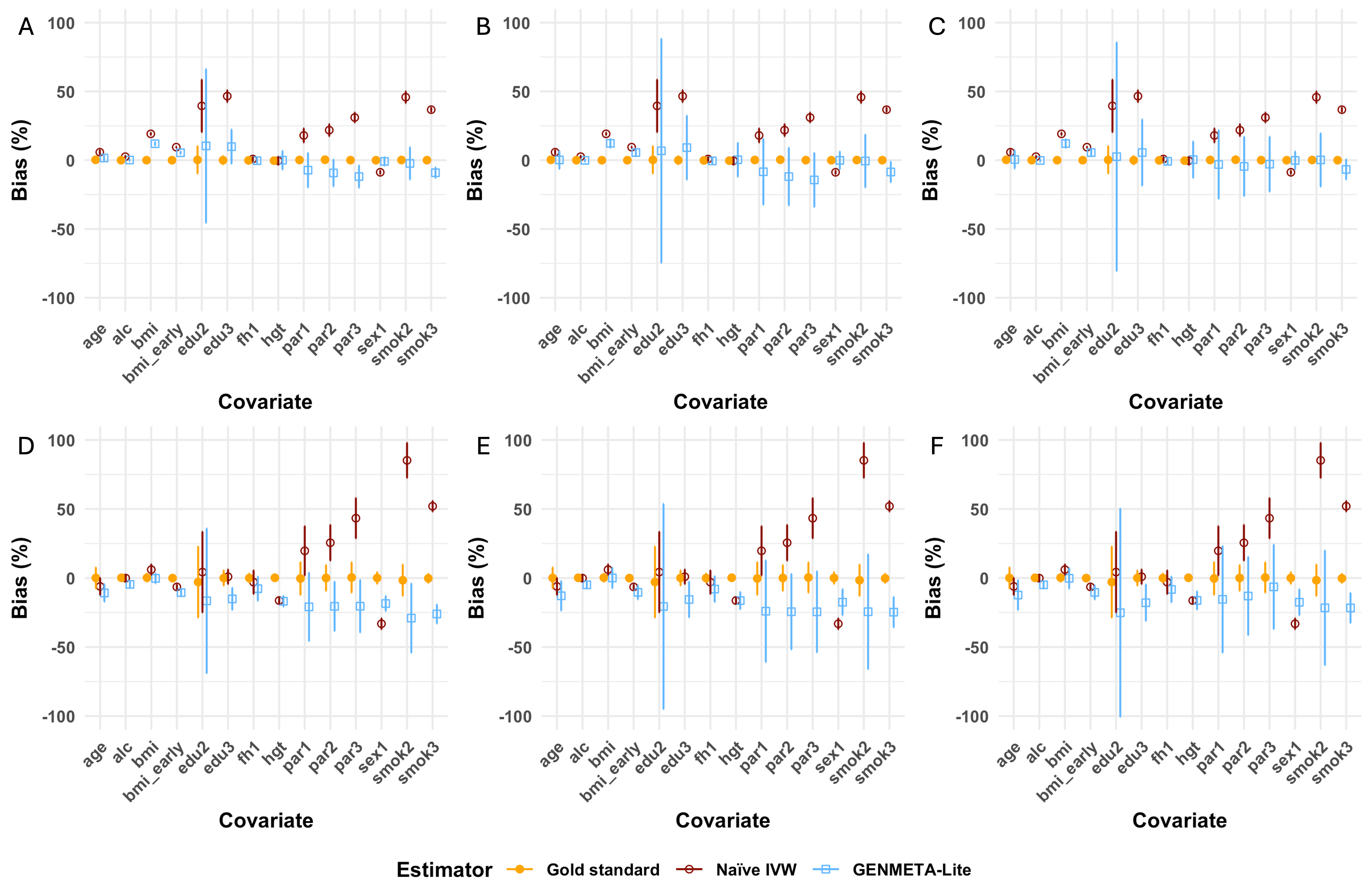}
    \caption{Results from Scenario 2, where underlying distributions differed between studies and the reference population. Both linear (A-C) and logistic (D-F) DGPs were used to simulate data. (A\&D) An unweighted reference population was used to estimate both $\widehat{\var}(\boldsymbol{X_s})$ and $\widehat{\var}(\boldsymbol{X})$. (B\&E) A weighted reference population was used to estimate both $\widehat{\var}(\boldsymbol{X_s})$ and $\widehat{\var}(\boldsymbol{X})$. (C\&F) $\widehat{\var}_s(\boldsymbol{X_s})$ was estimated in each study and used to build a partitioned covariance matrix, $\widehat{\var}_s(\boldsymbol{X})$, with a weighted reference population. Mean percent bias is shown for the gold standard estimator (filled gold circles), the na\"ive inverse-variance weighted meta-analysis estimator (open red circles), and GENMETA-Lite (open blue squares); error bars indicate empirical standard deviation of percent bias across simulations. Larger regression coefficients were used to simulate the outcome variables (see $\beta_2$ and $\textit{OR}_2$ in Table \ref{table:sup_sim_coeffs}). The reference population had sample size $n_{\text{ref}}=5000$.}
    \label{fig:diff_distrib_linear}
\end{figure}

When the reference population and each study had different underlying covariate distributions (violating Assumption~\ref{ass:same_distrib}), we compared three strategies for estimating the variance-covariance matrices used to construct $\boldsymbol{A_s}$ (Fig.~\ref{fig:diff_distrib_linear}; Table~\ref{table:supp_scenario_results}). First, using an unweighted reference population introduced noticeable bias (mean absolute bias: 5.8\% linear, 15.6\% logistic). Second, reweighting the reference population to balance moments and covariances with each study did not reduce this bias and substantially increased variance. Third, by combining study-specific covariance estimates with a weighted reference population to produce a partitioned variance-covariance matrix, the GENMETA-Lite estimator performed best of the three approaches, reducing bias (mean absolute bias: 3.2\% linear, 13.6\% logistic) while variance remained elevated relative to Scenario 1. Across all three approaches, GENMETA-Lite's bias remained lower than the na\"ive IVW estimator's (mean absolute bias: 20.5\% linear, 21.6\% logistic), and the gold standard estimator again showed negligible bias and small variance.

\subsubsection{Scenario 3: Coarsened covariate}
A linear constraint effectively handled a coarsened covariate (\textit{age}, binned in two of five reduced models): bias and variance for GENMETA-Lite were on par with Scenario 1 (mean absolute bias: 1.2\% linear, 16.0\% logistic; mean SD: 10.5\% linear, 14.7\% logistic; Fig.~\ref{fig:coarsened}; Table~\ref{table:supp_scenario_results}). Sensitivity analyses revealed that removing the coarsened variable from the final model reduced the empirical variance of the uncoarsened \textit{age} estimate more than three-fold, with a modest reduction in bias as well (Fig.~\ref{fig:coarsened_incl_vs_omitted_bias_sd}). The na\"ive IVW estimator showed consistently higher bias than GENMETA-Lite regardless of whether coarsened estimates were included or excluded from the meta-analysis (16.4\% and 18.0\% linear, 16.8\% and 16.1\% logistic, respectively).

\subsubsection{Scenario 4: Model selection}
\begin{figure}[!htb]
    \centering
    \includegraphics[height=0.6\textwidth]{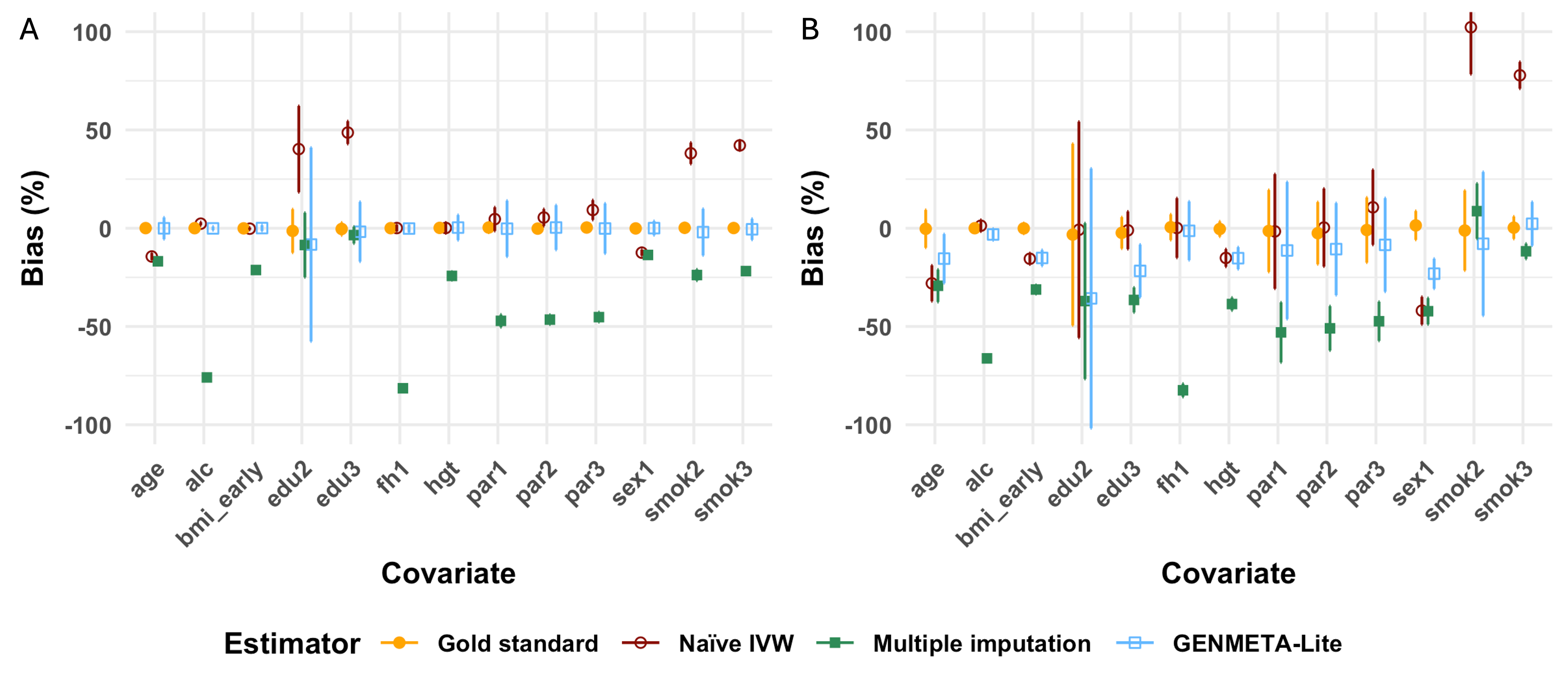}
    \caption{Results from Scenario 4 when model selection was used to omit \textit{BMI} from the final model, although \textit{BMI} had been included in three out of five reduced models, despite it  being excluded from the data generating process (DGP). Mean percent bias under (A) a linear DGP and (B) a logistic DGP are shown for the gold standard estimator (filled gold circles), the na\"ive inverse-variance weighted (IVW) meta-analysis estimator (open red circles), multiple imputation with parametric models (M=5; filled green squares), and GENMETA-Lite (open blue squares); error bars indicate empirical standard deviation of percent bias across simulations. Larger regression coefficients were used to simulate the outcome variables (see $\beta_2$ and $\textit{OR}_2$ in Table \ref{table:sup_sim_coeffs}). A partitioned covariance matrix, which combined covariance estimates from each study and covariance estimates from an unweighted reference population ($n_{\text{ref}} = 5000$), was used to estimate covariance across all covariates.}
    \label{fig:model_selection}
\end{figure}

When a linear constraint was used to remove \textit{BMI} from the final model, despite its inclusion in three of five reduced models and its absence from the DGP, bias for GENMETA-Lite decreased slightly relative to Scenario 1, with a modest increase in variance (mean absolute bias: 1.1\% linear, 13.2\% logistic; mean SD: 10.8\% linear, 19.8\% logistic; Fig.~\ref{fig:model_selection}; Table~\ref{table:supp_scenario_results}). Sensitivity analyses confirmed that including versus excluding \textit{BMI} did not meaningfully affect estimates for variables highly correlated with it (\textit{age}, \textit{early BMI}; Fig.~\ref{fig:model_sel_incl_vs_omitted_bias_sd}). By contrast, both the na\"ive IVW and imputation-based estimators showed increased bias and variance relative to Scenario 1, with the na\"ive IVW estimator showing especially large bias increases for sparse categorical levels under the logistic DGP (e.g., bias for \textit{smok2} exceeded 100\%).

\section{Application: Breast cancer risk prediction} \label{sec:RWD}
\subsection{Overview}
We applied our method to real world data to demonstrate its practicality. We used data from five cohorts to build a model for breast cancer risk in post-menopausal women (aged $\ge50$ years) using the risk factors and their categorizations from an existing model (iCARE-Lit;  \citealt{Choudhury2020_iCARE_Lit_model}). The risk factors are family history of breast cancer in first degree relatives, age at menarche, the number of full term pregnancies (``parity''), age at first birth, age at menopause, personal history of benign breast disease (BBD), height, alcohol consumption (drinks/week), smoking status (ever/never), and an interaction between menopausal hormone therapy, previously known as hormone replacement therapy (HRT; never user, former user, current estrogen and progesterone user, or current estrogen-only user) and body mass index (BMI). 

We compared the coefficient estimates obtained using GENMETA-Lite to the iCARE-Lit model coefficients. The iCARE-Lit coefficients were obtained from various published studies that estimated individual risk factor effect sizes. A concern with this approach is that the estimates for individual factors may not have accounted for all of the other factors going into the final iCARE-Lit model, so there is potential for miscalibration. In fact, validations in large prospective cohorts have shown that the iCARE-Lit model tends to overestimate risk in high risk participants \citep{brantleyPerformanceGeneralpopulationBreast2026,hursonProspectiveEvaluationBreastcancer2021}. Thus, we do not expect perfect agreement between the GENMETA-Lite estimates and the iCARE-Lit coefficients. 

Due to these limitations with the iCARE-Lit coefficients, we also compared the GENMETA-Lite estimates to estimates obtained from regression with the set of complete cases in the WHI cohort. Because WHI is the largest of the five cohorts (number of complete observations = 75,343, including 7,027 cases) and is only missing one risk factor (BBD), we expect the most unbiased and precise estimates from this single regression model. 

\subsection{Data}
\begin{figure}[!htb]
    \centering
    \includegraphics[height=0.6\textwidth]{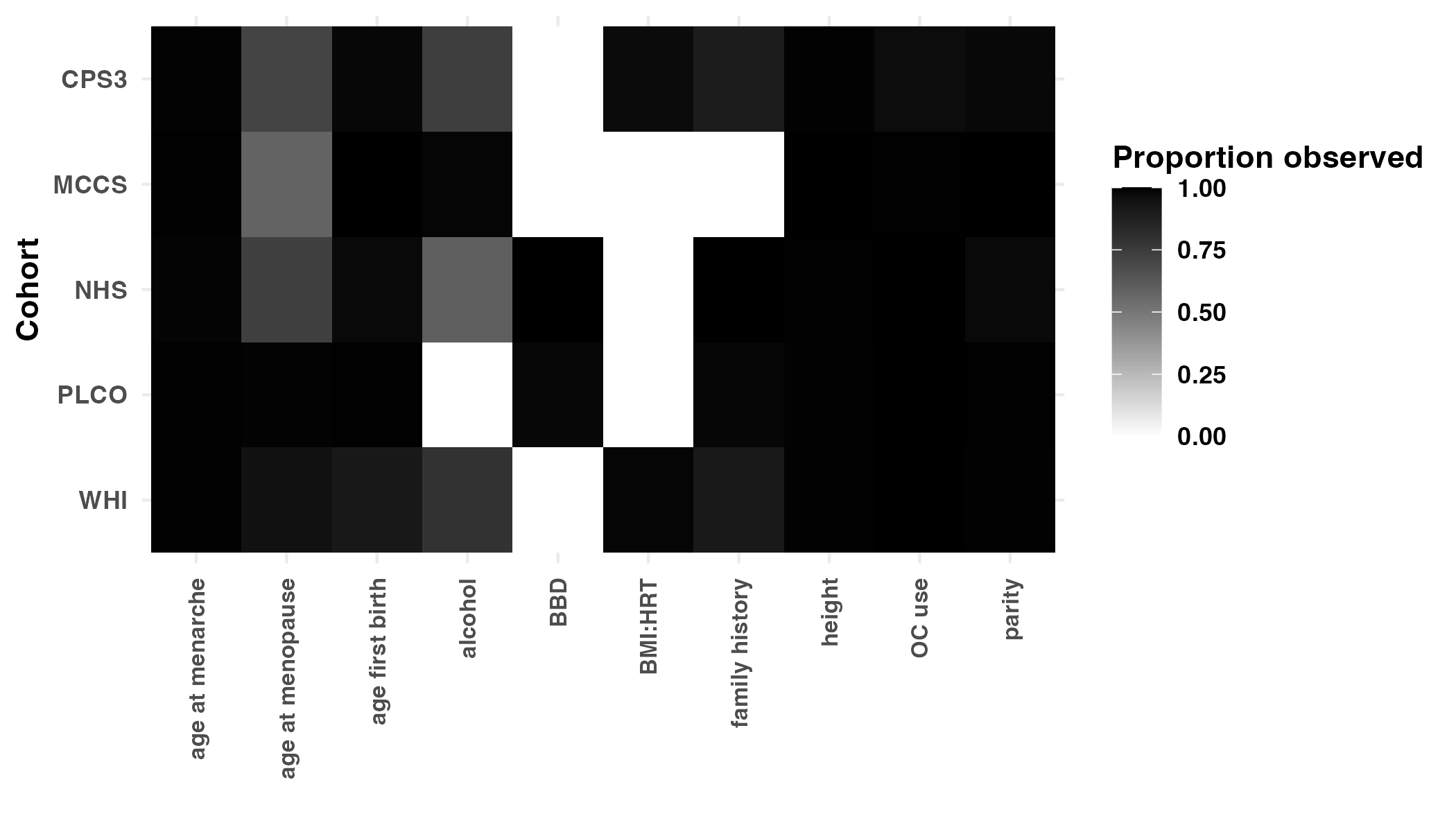}
    \caption{The proportion of participants in each cohort with observed data for covariates included in the iCARE-Lit breast cancer risk prediction model, accounting for structural missingness. Black = fully observed; white = missing by design. CPS3 = Cancer Prevention Study-3; MCCS = Melbourne Collaborative Cohort Study; NHS = Nurses' Health Study; PLCO = Prostate, Lung, Colorectal, and Ovarian Cancer Screening Trial; WHI = Women's Health Initiative; BBD = benign breast disease; BMI = body mass index; HRT = hormone replacement therapy; OC = oral contraceptive.}
    \label{fig:RWD_missingness}
\end{figure}

We used data from five prospective cohorts participating in the BCRPP, restricting our analyses to non-Hispanic White women $\ge50$ years old with no personal history of breast cancer. Breast cancer (BC) cases were defined as a diagnosis of invasive BC after enrollment; covariates were measured at enrollment. The following cohorts were chosen because they presented several realistic data challenges including a wide range in the number of participants, covariate distributions, and complex missingness patterns within and across cohorts (Fig.~\ref{fig:RWD_missingness}): 1) Cancer Prevention Study-3 (CPS3; participants = 90,514, BC cases = 1,313), 2) Melbourne Collaborative Cohort Study (MCCS; participants = 11,365, BC cases = 1,220), 3) Nurses' Health Study (NHS; participants = 24,771, BC cases = 2,870), 4) Prostate, Lung, Colorectal, and Ovarian Cancer Screening Trial (PLCO; participants = 67,353, BC cases = 5,409), and 5) Women's Health Initiative (WHI; participants = 119,848, BC cases = 10,742). 

The reference population, which included all covariates in the full model, was generated to be representative of non-Hispanic White women in the U.S. aged $\ge50$ with no prior history of cancer. To obtain the full set of covariates, we combined publicly available data from the National Health and Nutrition Examination Survey (NHANES), the National Health Interview Survey (NHIS), and the Breast Cancer Surveillance Consortium (BCSC). A full description of how the reference population was developed is available in the Supplementary Materials (Section~\ref{sec:ref_pop}). 

\subsection{Implementing GENMETA-Lite}
GENMETA-Lite was designed to be used in situations when covariates are missing by design, but it may also be used when data are missing at random (MAR). Here we demonstrate how to apply GENMETA-Lite sequentially to handle complex patterns of missingness (i.e., structural and MAR) observed both within and across cohorts (Fig.~\ref{fig:RWD_missingness}). First, we assessed missingness in each cohort and identified dominant missingness patterns, defined as groups of $\ge10\%$ of participants that had the same missing covariates (Fig.~\ref{fig:id_dominant_missingness_patterns}). If there were more than four dominant missingness patterns present, we combined less frequent patterns until we obtained a maximum of four patterns per cohort that together comprised $\ge90\%$ of the cohort. This strategy limited the removal of individuals with missing data to  $\le10\%$. We then fit a reduced logistic regression model within each subset of participants defined by a given missingness pattern (henceforth referred to as a ``sub-study''), additionally adjusting for participant age, with diagnosis of breast cancer as the outcome. To combine estimates within each cohort, we applied GENMETA-Lite using the complete cases within the cohort as the ``reference population.'' We estimated $\widehat{\var}_s(\boldsymbol{X_s})$ within each sub-study and used that to build a partitioned covariance matrix for $\widehat{\var}(\boldsymbol{X})$ with the complete cases weighted to match the first moment for the binned \textit{age} covariate within each sub-study. 

We then applied GENMETA-Lite again to combine the estimates across cohorts. Robust sandwich variance estimates were used unless there were fewer than 100 observations in a given level, in which case model-based variance estimates were used. For each cohort, we used the complete cases to estimate $\widehat{\var}_s(\boldsymbol{X_s})$ and used this to build a partitioned covariance matrix for $\widehat{\var}_s(\boldsymbol{X})$ with the reference population weighted to balance the first two moments, including the covariance, for \textit{age} and \textit{BMI}.

\subsection{Results}
Due to complex missingness patterns within cohorts, we  fit four reduced models for CPS3 and NHS, three for WHI, and two for MCCS and PLCO (Fig.~\ref{fig:nested_model_missingness}). When these estimates were combined within cohorts using GENMETA-Lite, the estimates were generally similar across cohorts, though standard errors varied based on sample sizes (Fig.~\ref{fig:RWD_est_across_cohorts}).

\begin{figure}[!htb]
    \centering
    \includegraphics[height=0.6\textwidth]{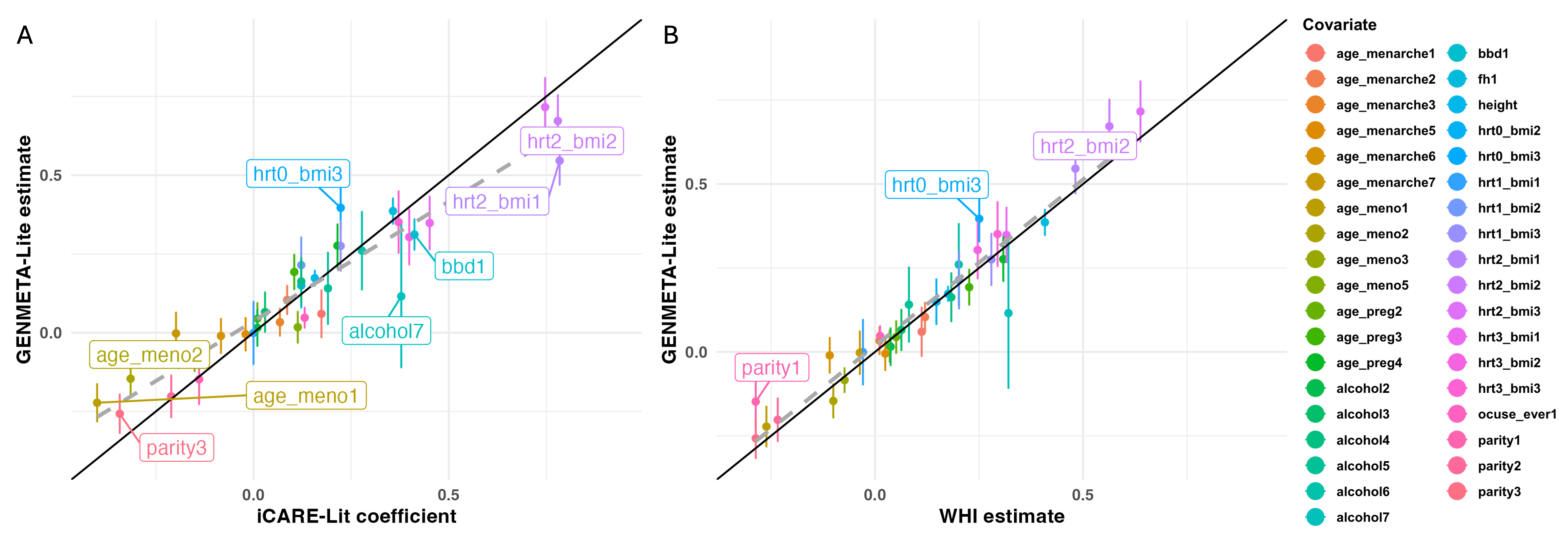}
    \caption{Risk factor point estimates and 95\% confidence intervals generated using the GENMETA-Lite method compared to (A) literature-based iCARE-Lit breast cancer risk model coefficients, and (B) estimates obtained from fitting the model with complete cases from the WHI cohort. The covariance matrix for GENMETA-Lite was estimated as a partitioned matrix that combined covariance estimates from each cohort and a reweighted reference population. Estimates whose confidence intervals do not cross the $y=x$ (solid) line are labeled. The dashed line represents the least-squares regression line. BBD = benign breast disease; BMI = body mass index; HRT = hormone replacement therapy; OC = oral contraceptive; age\_meno = age at menopause.}
    \label{fig:RWD_combined_opt1}
\end{figure}

The GENMETA-Lite estimates were strongly correlated with the literature-based iCARE-Lit estimates (Pearson's correlation = 0.94; Fig.~\ref{fig:RWD_combined_opt1}A). However, the larger magnitude coefficients from the iCARE-Lit model tended to be attenuated toward the null when estimated using GENMETA-Lite, and the slope of the least-squares regression line was significantly less than one (slope = 0.76, 95\% CI: 0.67 to 0.85; $p = 6.1 \times 10^{-6}$). The estimates that differed the most between GENMETA-Lite and iCARE-Lit were interactions between HRT and BMI, and main effects for BBD, age at menopause, the highest category for parity, and the highest category for alcohol consumption (Fig.~\ref{fig:RWD_combined_opt1}A).

There was stronger agreement between the GENMETA-Lite estimates and the estimates obtained from regression of the complete cases in the WHI cohort (Pearson's correlation = 0.94; Fig.~\ref{fig:RWD_combined_opt1}B), and the slope of the least-squares regression line was not significantly different than one (slope = 0.95, 95\% CI: 0.84 to 1.06; p = 0.83). Most of the risk factor estimates were very similar between GENMETA-Lite and WHI. Despite having observed some bias toward the null in the simulation scenarios for logistic regression settings (e.g., Fig.~\ref{fig:basecase_logistic_parametric_partitioned_var}B), the strong agreement between the GENMETA-Lite and WHI coefficient estimates here indicate that GENMETA-Lite can produce unbiased parameter estimates for all risk factors under a full model. The underestimation of effects of the GENMETA-Lite estimates relative to the iCARE-Lit coefficients is likely driven by miscalibration of the iCARE-Lit model.

\section{Discussion and conclusion} \label{sec-conc}

In summary, we propose GENMETA-Lite, a linear version of the GENMETA method, as a practical and scalable approach for conducting multivariate meta-analysis of model parameter estimates across different studies with heterogeneous covariate data. GENMETA-Lite is a non-iterative approach that maintains the principles of federated learning by not requiring access to individual-level data. We propose additional functionalities for handling covariate shift using entropy balancing weights and partitioned covariance matrices, as well as parameter constraining for handling coarsened covariates and conducting model selection. Our extensive simulation studies with realistic data complexities showed good bias control for GENMETA-Lite relative to other off-the-shelf methods, such as na\"ive IVW or multiple imputation estimators. Application of the method to large scale data from the BCRPP further demonstrated how GENMETA-Lite can be applied sequentially to handle data missing at random (within studies) and missing by design (across studies). In short, GENMETA-Lite is a practical and valid method for producing model parameter estimates in complex analysis settings.

While GENMETA provided the first theoretical foundation for meta-analysis in the non-linear heterogeneous data setting \citep{Kundu2019}, the algorithm requires iterative fitting and can have problems converging due to the non-convex nature of the optimization function. Thus, it may not currently be practical for complex data analyses that require fitting many different models. Additionally, GENMETA requires not only parameters for risk factors of interest, but also effect estimates from study-specific confounding factors such as intercept terms, age, study center, etc. Such information may be shared in a collaboration setting but may not be available if information is gleaned from the published literature. In contrast, GENMETA-Lite is a one-shot estimator that only requires information on the key covariates of interest (i.e. parameter estimates and covariance matrices), making it quite practical.

The current limitations of the GENMETA-Lite estimator help identify future research directions. GENMETA-Lite assumes a linear DGP. Through simulation studies, we showed that while GENMETA-Lite provides much better bias control compared to alternative methods, it can have non-negligible bias in a logistic regression setting. One path to reduce the bias for non-linear regression would be to use the GENMETA-Lite estimates as initialization values in GENMETA and replace the underlying variance-covariance matrix $\boldsymbol{V}$ with the corresponding matrix from the linear case while preserving the original estimating functions for GENMETA. Additional studies are also needed to investigate performance of GENMETA-Lite for other non-linear DGPs such as log linear, probit, and Cox proportional hazards models. Preliminary studies suggest residual bias is the worst in the case of logistic regression (results to be presented elsewhere). Finally, GENMETA-Lite assumes fixed effects, implying the same effect sizes underlie each study, but it could be extended to account for random effects \citep{dersimonianMetaanalysisClinicalTrials1986,higginsReevaluationRandomeffectsMetaanalysis2009,linRelativeEfficiencyUsing2010}. 

In conclusion, GENMETA-Lite is a robust and practical estimator for conducting multivariate meta-analysis across studies with disparate covariate information. This one-step estimator does not require sharing of individual-level data, and it is computationally fast and stable, making it tractable to conduct complex data analyses in a federating setting. We anticipate popularity of our R package given the meta-analysis challenges we tackle are very common in large consortium settings.

\section{Acknowledgements}\label{acknowledgements}

The authors gratefully acknowledge NIH Grants R01HG013137, U01CA249866, R01CA228198, NCI Intramural Research, and the Fund for Innovation in Cancer Informatics Foundation, for funding this research. The American Cancer Society funds the creation, maintenance, and updating of the Cancer Prevention Study-3 cohort. The authors express sincere appreciation to all Cancer Prevention Study-3 participants, and to each member of the study and biospecimen management group. The authors would like to acknowledge the contribution to this study from central cancer registries supported through the Centers for Disease Control and Prevention's National Program of Cancer Registries and cancer registries supported by the National Cancer Institute's Surveillance Epidemiology and End Results Program. We also thank Prosenjit Kundu for his thoughtful contributions which improved the manuscript.

 \section{Disclosure statement}\label{disclosure-statement}

 The authors have no conflicts of interest to report.
 
 The authors disclose that generative AI (Claude version 1.18286.0) was used to assist with developing code for the variance estimator and for proofreading the final manuscript. 

 \section{Data Availability Statement}\label{data-availability-statement}

The cohort data that support the findings of this study are available through the Breast Cancer Risk Prediction Project (https://epidataplatforms.cancer.gov/bcrpp/). Restrictions apply to the availability of these data, which were used under license for this study. The reference population data, derived from publicly available sources, are available from the corresponding author upon reasonable request. 

  \putbib          
\end{bibunit}

 \phantomsection\label{supplementary-material}


\renewcommand{\thefigure}{S\arabic{figure}}
\renewcommand{\thetable}{S\arabic{table}}
\renewcommand{\thesection}{S\arabic{section}}
\renewcommand{\theequation}{S\arabic{equation}}
\renewcommand{\theHfigure}{S\arabic{figure}}
\renewcommand{\theHtable}{S\arabic{table}}
\renewcommand{\theHsection}{S\arabic{section}}
\renewcommand{\theHequation}{S\arabic{equation}}
\makeatother

\title{\Large \bf{Supplementary Materials}}
\author{}
\maketitle


\begin{bibunit}

\def\spacingset#1{\renewcommand{\baselinestretch}%
{#1}\small\normalsize} \spacingset{1}
\setcounter{table}{0}
\setcounter{figure}{0}
\setcounter{equation}{0}


\section{Methods}

\subsection{Variance of the GENMETA-Lite estimator} \label{sec:var_estimator_deriv}

$\boldsymbol{\bG}$ solves the estimating equation $\sum_s \boldsymbol{A_s}\T \boldsymbol{V_s}^{-1}(\ths-\boldsymbol{A_s\beta})=\boldsymbol{0}$, and is a smooth function of two independent groups of inputs: the reduced-model estimates $\ths$ (from the
$n_s$ study-$s$ observations) and the estimated covariance matrices that define
$\boldsymbol{A_s}=\vh(\boldsymbol{X_s})^{-1}\boldsymbol{\Delta_s}\vh(\boldsymbol{X})$ (from the $n_{\mathrm{ref}}$ reference observations and,
under the partitioned option, the studies). A first-order (delta-method) expansion of the
estimating equation about the truth, using $-(\partial\boldsymbol{\Psi}/\partial\boldsymbol{\beta})^{-1}=\M$ and the fact
that the residual $\boldsymbol{\theta_s}-\boldsymbol{A_s\beta}$ vanishes at the truth, gives
\begin{equation}
\bG-\boldsymbol{\beta}=\M\sum_s \boldsymbol{A_s}\T \boldsymbol{V_s}^{-1}(\ths-\boldsymbol{\theta_s})
\;-\;\M\sum_s \boldsymbol{A_s}\T \boldsymbol{V_s}^{-1}\,(d\boldsymbol{A_s})\,\boldsymbol{\beta}\,
\label{eq:S71-exp}
\end{equation}
where $d\boldsymbol{A_s}$ is the perturbation of $\boldsymbol{A_s}$ induced by the estimated covariance matrices $d\boldsymbol{\Sigma}:=\widehat{\boldsymbol{\Sigma}}-\boldsymbol{\Sigma}$ (equivalently $\vecop\widehat{\boldsymbol{\Sigma}}-\vecop\boldsymbol{\Sigma}$).
Differentiating $\boldsymbol{A_s}=\vh(\boldsymbol{X_s})^{-1}\boldsymbol{\Delta_s}\vh(\boldsymbol{X})$,
\begin{equation}
d\boldsymbol{A_s}=\vh(\boldsymbol{X_s})^{-1}\,\boldsymbol{\Delta_s}\,(d\boldsymbol{\Sigma})\,\big(\boldsymbol{\Ip}-\boldsymbol{\Delta_s}\T \boldsymbol{A_s}\big),\qquad \boldsymbol{\Sigma}:=\vh(\boldsymbol{X}).
\label{eq:S71-dA}
\end{equation}
Because the studies and the reference population are independent samples, the two right-hand terms of
\eqref{eq:S71-exp} are uncorrelated and their variances add. The first gives the efficient
meta-analysis variance
\[
\var\!\Big(\M\textstyle\sum_s \boldsymbol{A_s}\T \boldsymbol{V_s}^{-1}(\ths-\boldsymbol{\theta_s})\Big)
=\M\Big(\sum_s \boldsymbol{A_s}\T \boldsymbol{V_s}^{-1}\boldsymbol{V_s V_s}^{-1}\boldsymbol{A_s}\Big)\M=\M ,
\]
i.e.\ Term $\M$ in \eqref{eq:GENMETA-Lite_var}. The second term is the variance associated with estimating the covariance matrices $\widehat{\boldsymbol{\var}}_{\boldsymbol{\hat{A}}_s}$, treated separately for the two covariance options below.

\medskip
\textbf{(a) Reference-only covariance ($\widehat{\boldsymbol{\var}}_{\boldsymbol{\hat{A}}_s} = \boldsymbol{\var_{\mathrm{ref}}}$).}
Here $\boldsymbol{\Sigma}=\vh(\boldsymbol{X})$ is estimated from the reference population as
$\widehat{\boldsymbol{\Sigma}}=\tfrac1{n_{\mathrm{ref}}}\sum_{j}(\boldsymbol{x_j}-\bar{\boldsymbol{x}})(\boldsymbol{x_j}-\bar{\boldsymbol{x}})\T$, whose per-observation
influence function is $(\boldsymbol{x_j}-\bar{\boldsymbol{x}})(\boldsymbol{x_j}-\bar{\boldsymbol{x}})\T-\widehat{\boldsymbol{\Sigma}}$. Substituting into
\eqref{eq:S71-dA}--\eqref{eq:S71-exp}, the reference influence function of $\bG$ is
\begin{equation}
\boldsymbol{\mathrm{IF}_j}=-\,\M\sum_s \boldsymbol{A_s}\T \boldsymbol{V_s}^{-1}\,\vh(\boldsymbol{X_s})^{-1}\boldsymbol{\Delta_s}
\Big[(\boldsymbol{x_j}-\bar{\boldsymbol{x}})(\boldsymbol{x_j}-\bar{\boldsymbol{x}})\T-\widehat{\boldsymbol{\Sigma}}\Big]\big(\boldsymbol{\Ip}-\boldsymbol{\Delta_s}\T \boldsymbol{A_s}\big)\,\bG ,
\label{eq:S71-IF}
\end{equation}
and, because the reference population contribution to $\bG-\boldsymbol{\beta}$ is $\tfrac1{n_{\mathrm{ref}}}\sum_j \boldsymbol{\mathrm{IF}_j}$,
\begin{equation}
\widehat{\boldsymbol{\var}}_{\mathrm{ref}}
=\frac{1}{n_{\mathrm{ref}}^2}\sum_{j=1}^{n_{\mathrm{ref}}}
\big(\boldsymbol{\mathrm{IF}_j}-\overline{\boldsymbol{\mathrm{IF}}}\,\big)\big(\boldsymbol{\mathrm{IF}_j}-\overline{\boldsymbol{\mathrm{IF}}}\,\big)\T ,
\qquad \overline{\boldsymbol{\mathrm{IF}}}=\tfrac1{n_{\mathrm{ref}}}\textstyle\sum_j\boldsymbol{\mathrm{IF}_j}
\label{eq:S71-Vref}
\end{equation}
This is $O(1/n_{\mathrm{ref}})$ and proportional to the signal $\boldsymbol{\beta}$ (through
\eqref{eq:S71-IF}); it is negligible only when $n_{\mathrm{ref}}\gg n_s$ or $\boldsymbol{\beta}$ is null.
If the reference is reweighted (weights $w_j$; Section~\ref{sec:cov_shift}), replace \eqref{eq:S71-Vref} by
the effective-sample-size form $\widehat{\boldsymbol{\var}}_{\mathrm{ref}}
=\big(\sum_j w_j\big)^{-2}\sum_j w_j^{2}
\big(\boldsymbol{\mathrm{IF}_j}-\overline{\boldsymbol{\mathrm{IF}}}_w\,\big)\big(\boldsymbol{\mathrm{IF}_j}-\overline{\boldsymbol{\mathrm{IF}}}_w\,\big)\T$
with $\overline{\boldsymbol{\mathrm{IF}}}_w$ the weighted mean and $\bar{\boldsymbol{x}}$ the weighted mean in
\eqref{eq:S71-IF}--\eqref{eq:S71-Vref} (this treats the weights as fixed; see Note below).

\medskip
\textbf{(b) Partitioned covariance ($\widehat{\boldsymbol{\var}}_{\boldsymbol{\hat{A}}_s} = \boldsymbol{\var_{\mathrm{ref}}}+\boldsymbol{\var_{\mathrm{study}}}$).}
With the partitioned matrix, $\boldsymbol{A_s}=\big[\,\boldsymbol{I_{k_s}}\ \ \boldsymbol{B_s}\,\big]$ with
$\boldsymbol{B_s}=\vh_s(\boldsymbol{X_s})^{-1}\ch(\boldsymbol{X_s},\Xt)$: the observed block is the identity, the study's own
$\vh_s(\boldsymbol{X_s})$ inverts the block, and the reference enters only through the cross-covariance
$\ch(\boldsymbol{X_s},\Xt)$.

\emph{Reference part.} $\boldsymbol{\var_{\mathrm{ref}}}$ is computed exactly as in \eqref{eq:S71-IF}--\eqref{eq:S71-Vref},
except that in \eqref{eq:S71-IF} the influence function of the covariance is restricted to its
cross block: only $\ch(\boldsymbol{X_s},\Xt)$ is reference-estimated (the within-study block $\vh(\boldsymbol{X_s})$ is a
study quantity, and $\vh(\Xt)$ does not enter $\boldsymbol{A_s}$), so replace
$[(\boldsymbol{x_j}-\bar{\boldsymbol{x}})(\boldsymbol{x_j}-\bar{\boldsymbol{x}})\T-\widehat{\boldsymbol{\Sigma}}]$ by its $(\boldsymbol{X_s},\Xt)$ off-diagonal block (the
within-$\boldsymbol{X_s}$ entries set to zero).

\emph{Study part.} Estimating $\vh(\boldsymbol{X_s})$ from the study makes $\boldsymbol{A_s}$ depend on the
same sample as $\ths$, so the two influences combine. For study $s$ let $\boldsymbol{\beta}_{\Xt}$ be the
sub-vector of $\bG$ for the covariates $\Xt$ missing from the study,
$\boldsymbol{B_s}=\vh(\boldsymbol{X_s})^{-1}\ch(\boldsymbol{X_s},\Xt)$, $g_s=B_s\,\boldsymbol{\beta}_{\Xt}$, and
$\widehat{\epsilon}_{s,i}=y_{s,i}-\boldsymbol{x_{s,i}}\T\ths$ the reduced-model residuals ($\boldsymbol{x_{s,i}}$
mean-centred). Define the per-observation $p$-vectors
\begin{align*}
\boldsymbol{c_{s,i}}&=\boldsymbol{A_s}\T \boldsymbol{V_s}^{-1}\vh(\boldsymbol{X_s})^{-1}\,\boldsymbol{x_{s,i}}\,\widehat\epsilon_{s,i},\\
\boldsymbol{d_{s,i}}&=\boldsymbol{A_s}\T \boldsymbol{V_s}^{-1}\vh(\boldsymbol{X_s})^{-1}\big(\boldsymbol{x_{s,i}}\boldsymbol{x_{s,i}}\T-\vh(\boldsymbol{X_s})\big)\,\boldsymbol{g_s} .
\end{align*}
Then
\begin{equation}
\widehat{\boldsymbol{\var}}_{\mathrm{study}}
=\sum_{s=1}^S \frac{1}{n_s}\,\M\Big[\widehat{\cov}(\boldsymbol{d_s})
+\widehat{\cov}(\boldsymbol{c_s},\boldsymbol{d_s})+\widehat{\cov}(\boldsymbol{d_s},\boldsymbol{c_s})\Big]\M 
\label{eq:S71-Vstudy}
\end{equation}
where $\widehat{\cov}(\cdot)$ and $\widehat{\cov}(\cdot,\cdot)$ are the sample (cross-)covariances
over $i=1,\dots,n_s$. Equation~\eqref{eq:S71-Vstudy} is $O(1/n_s)$ and proportional to
$\boldsymbol{\beta}_{\Xt}\boldsymbol{\beta}_{\Xt}\T$: it is $0$ when the study observes all covariates ($g_s$ empty) or
$\boldsymbol{\beta}_{\Xt}=\boldsymbol{0}$, and the cross terms $\widehat{\cov}(\boldsymbol{c_s},\boldsymbol{d_s})$ vanish under homoskedastic,
symmetric errors (leaving only the fourth-moment $\widehat{\cov}(\boldsymbol{d_s})$ term). The full
partitioned variance is $\var(\bG)=\M+\boldsymbol{\var}_{\mathrm{ref}}+\boldsymbol{\var}_{\mathrm{study}}$.

\medskip
\textbf{(c) Constrained estimator.}
With $\bG^{c}=\boldsymbol{P}\bG$, $\boldsymbol{P}=\boldsymbol{I_p}-\M \boldsymbol{C}\T(\boldsymbol{C}\M \boldsymbol{C}\T)^{-1}\boldsymbol{C}$, and (under a correctly specified
constraint $\boldsymbol{C\beta}=\boldsymbol{0}$) $\boldsymbol{C}\bG=O_p(n^{-1/2})$, the derivative of $\bG^{c}$ through the estimated
$\M$ is higher order, so to first order $\boldsymbol{\mathrm{IF}}_{\beta^c}=\boldsymbol{P}\,\boldsymbol{\mathrm{IF}}_{\beta}$ and
\begin{equation}
\var\!\big(\bG^{c}\big)=\boldsymbol{P}\,\var(\bG)\,\boldsymbol{P}\T
=\boldsymbol{P}\big(\M+\boldsymbol{\var}_{\mathrm{ref}}+\boldsymbol{\var}_{\mathrm{study}}\big)\boldsymbol{P}\T 
\label{eq:S71-Vc}
\end{equation}

\medskip
\emph{Note.} Equations \eqref{eq:S71-Vref}--\eqref{eq:S71-Vstudy} treat any
covariate-shift weights as fixed. If the weights are estimated (e.g.\ entropy balancing), they
are additional estimated parameters whose influence function enters $\var_{\mathrm{ref}}$ as a
further additive term, and, because the balancing targets the study moments, couples the
reference and study samples. A joint resampling scheme (resampling reference and study rows and
re-estimating the weights) targets the full variance including this contribution.

\subsection{Entropy balancing} \label{sec:entropy_balancing}

Entropy balancing works by minimizing the distance between the original weights of the reference population (i.e., ``base weights''; base weights $= 1/n_{\text{ref}}$, if they are not explicitly defined) and estimated weights, subject to pre-defined ``balance constraints'' that determine which moments must be balanced for which variables. Additionally, the estimated weights must sum to unity (i.e., ``normalized weights''). The loss function, assuming an entropy divergence distance metric proposed by \citet{Kullback1959}, and enforcing the balance constraints and the normalized weighting with Lagrange multipliers, can be formally written as follows:
$$
\min_{\boldsymbol{W},\lambda_0,\boldsymbol{Z}} L^p = \sum_{\{i|D=0\}} w_i \log\left( \frac{w_i}{q_i} \right) + \sum_{r=1}^{R} \lambda_r \left(\sum_{\{i|D=0\}}w_i c_{ri}(X_i) - m_r \right) + (\lambda_0 - 1) \left( \sum_{\{i|D=0\}}w_i - 1 \right)
$$
where $D = 0$ indicates membership in the reference population for which weights will be estimated, $w_i$ is the estimated weight and $q_i$ is the base weight, $\boldsymbol{Z} = \{\lambda_1, ..., \lambda_R \}^\top$ is a vector of Lagrange multipliers for the $R$ balance constraints $c_{ri}(X_i)$ with moments $m_r$, and $\lambda_0 - 1$ is the Lagrange multiplier for the normalization constraint. See full details of implementation in \citet{Hainmueller2012}.

\subsection{Study-specific variables} \label{sec:study-spec-vars}

Suppose a reduced model includes study-specific variables. For example, a study may have collected data from participants at different centers, and they adjust for center when they fit their reduced model:
\begin{eqnarray*}
\boldsymbol{y} = \boldsymbol{D\alpha} + \boldsymbol{X_s} \tilde{\boldsymbol{\theta_s}} + \boldsymbol{\epsilon}
\end{eqnarray*}
where $\boldsymbol{D}$ is an $n_s$ x $d_s$ matrix of the study-specific (nuisance) variables (e.g. assessment center), $\boldsymbol{X_s}$ is a $n_s$ x $k_s$ matrix of variables of interest that will be included in the full model, and $\boldsymbol{\epsilon} \sim N(\boldsymbol{0},\sigma^2)$
is an error term. After model fitting using individual-level data, the cohort administrators will share $\hat{\tilde{\boldsymbol{\theta_s}}}$ and empirical
$\widehat{\var}(\boldsymbol{X_s})$. However, our linear estimator relies on
$\hat{\boldsymbol{\theta_s}} = [\boldsymbol{X_s}^T\boldsymbol{X_s}]^{-1}\boldsymbol{X_s}^T\boldsymbol{y}$. So how do
$\hat{\tilde{\boldsymbol{\theta_s}}}$ and $\hat{\boldsymbol{\theta_s}}$ relate, and can we derive an unbiased estimate of $\hat{\boldsymbol{\theta}}$ from what is provided to us?
Define $\hat{\boldsymbol{\gamma}} = \begin{bmatrix} \hat{\boldsymbol{\alpha}} & \hat{\tilde{\boldsymbol{\theta_s}}} \end{bmatrix}^\top = [\boldsymbol{Z}^T\boldsymbol{Z}]^{-1}\boldsymbol{Z}^T\boldsymbol{y}$
where $\boldsymbol{Z} = \begin{bmatrix} \boldsymbol{D} & \boldsymbol{X_s} \end{bmatrix}$. Then we can derive the following closed-form solution:
\begin{eqnarray} \label{eq:theta_tilde}
\hat{\tilde{\boldsymbol{\theta_s}}} &=& - \boldsymbol{W}\boldsymbol{X_s}^T\boldsymbol{D}[\boldsymbol{D}^T\boldsymbol{D}]^{-1}\boldsymbol{D}^T\boldsymbol{y} + \boldsymbol{W}\boldsymbol{X_s}^T\boldsymbol{y} \notag \\
&=& - \boldsymbol{W}\boldsymbol{X_s}^T\boldsymbol{H_D}\boldsymbol{y} + \boldsymbol{W}\boldsymbol{X_s}^T\boldsymbol{y} \label{theta_tilde}
\end{eqnarray}
where $\boldsymbol{H_D} = \boldsymbol{D}[\boldsymbol{D}^T\boldsymbol{D}]^{-1}\boldsymbol{D}^T$ and $\boldsymbol{W} \equiv [\boldsymbol{X_s}^T\boldsymbol{X_s} - \boldsymbol{X_s}^T\boldsymbol{H_D}\boldsymbol{X_s}]^{-1}$. We can further show that $\boldsymbol{W} = [\widetilde{\boldsymbol{X_s}}^T\widetilde{\boldsymbol{X_s}}]^{-1}$ where $\widetilde{\boldsymbol{X_s}}$
is the residual of the covariates of interest $\boldsymbol{X_s}$ regressed on the study-specific covariates $\boldsymbol{D}$, i.e. $\widetilde{\boldsymbol{X_s}} = \boldsymbol{X_s} - \widehat{\boldsymbol{X_s}}$:
\begin{eqnarray}
\boldsymbol{W} &=& [\boldsymbol{X_s}^T\boldsymbol{X_s} - \boldsymbol{X_s}^T\boldsymbol{H_D}\boldsymbol{X_s}]^{-1} \notag \\
&=& [\boldsymbol{X_s}^T(\boldsymbol{I} - \boldsymbol{H_D})\boldsymbol{X_s}]^{-1} \notag \\
&=&  [\boldsymbol{X_s}^T(\boldsymbol{I} - \boldsymbol{H_D})(\boldsymbol{I}-\boldsymbol{H_D})\boldsymbol{X_s}]^{-1} \because{\text{projection matrices are idempotent}} \notag \\
&=& [(\boldsymbol{X_s} - \widehat{\boldsymbol{X_s}})^T(\boldsymbol{X_s} - \widehat{\boldsymbol{X_s}})]^{-1} \notag  \\
&=& [\widetilde{\boldsymbol{X_s}}^T\widetilde{\boldsymbol{X_s}}]^{-1} \label{eq:W}
\end{eqnarray}
From \eqref{eq:theta_tilde} and \eqref{eq:W}, we can simplify $\hat{\tilde{\boldsymbol{\theta_s}}}$ as follows:
\begin{eqnarray*}
\hat{\tilde{\boldsymbol{\theta_s}}} &=& - \boldsymbol{W}\boldsymbol{X_s}^T\boldsymbol{H_D}\boldsymbol{y} + \boldsymbol{W}\boldsymbol{X_s}^T\boldsymbol{y} \\
&=& \boldsymbol{W}\boldsymbol{X_s}^T(\boldsymbol{I} - \boldsymbol{H_D})\boldsymbol{y} \\
&=& [\widetilde{\boldsymbol{X_s}}^T\widetilde{\boldsymbol{X_s}}]^{-1}\widetilde{\boldsymbol{X_s}}^T\boldsymbol{y} 
\end{eqnarray*}
Note that if $\boldsymbol{X_s} \perp \boldsymbol{D}$, then $\boldsymbol{X_s}^T\boldsymbol{D} = \boldsymbol{0} \implies \boldsymbol{W} = [\boldsymbol{X_s}^T\boldsymbol{X_s}]^{-1}\implies \hat{\tilde{\boldsymbol{\theta_s}}} = [\boldsymbol{X_s}^T\boldsymbol{X_s}]^{-1}\boldsymbol{X_s}^T\boldsymbol{y} = \hat{\boldsymbol{\theta_s}}$. However, as $\operatorname{cov}(\boldsymbol{X_s},\boldsymbol{D})$ increases, so will the distance between
$\hat{\tilde{\boldsymbol{\theta_s}}}$ and the desired $\hat{\boldsymbol{\theta_s}}$.

\section{Simulation studies}

\subsection{Estimators for comparison} \label{sec:comp_estimators}  
\subsubsection{Gold standard estimator}
After the design matrix and outcome variable were generated, but prior to imposing missingness by design, full regression models were fit within each study. Those estimates were combined using an IVW estimator to obtain the ``gold standard'' estimates. While this estimator cannot be calculated in practice, it helps identify any intrinsic biases or limitations in our simulations.

\subsubsection{Na\"ive inverse-variance weighted estimator}
After fitting reduced models in each study, the ``na\"ive IVW'' estimator was generated by running IVW meta-analysis for each covariate separately, only averaging across the estimates available from the subset of reduced models $S_q$ in which the $q$th covariate was included:

\[\boldsymbol{\widehat{\beta_{IVW_q}}} = \left(\sum_{s=1}^{S_q}\boldsymbol{V}_{s_q}^{-1}\right)^{-1} \sum_{s=1}^{S_q} \boldsymbol{V}_{s_q}^{-1} \boldsymbol{\widehat{\theta_{s_q}}}\]

where \(\boldsymbol{V}_{s_q}\) is the variance of the $q$th parameter estimate for the \(s\)th study (\(\boldsymbol{\widehat{\theta_{s_q}}}\)). The variance of the estimator is:

\[var(\boldsymbol{\widehat{\beta_{IVW_q}}}) = \left( \sum_{s=1}^{S_q} \boldsymbol{V}_{s_q}^{-1} \right)^{-1}\].

This estimator is expected to be biased unless the structurally missing covariates are independent of the observed covariates, which is unlikely in many real-world applications (and does not hold in our simulations). If a coarsened covariate was present (Scenario 3), reduced models were fit with the coarsened variable (\textit{age\_bin}), and na\"ive IVW estimates for the \textit{age} variable were calculated both including and excluding coefficient estimates for the coarsened variable.

\subsubsection{Imputation-based estimator}
In practice, when individual-level data are available, missingness may be addressed prior to analyses using multiple imputation \citep{Rubin1987}. For parametric imputation, imputation formulae were defined using the set of covariates known to be correlated with each missing variable (see Section~\ref{sec:dgp}). We assumed Gaussian distributions for continuous variables, binomial distributions for binary variables, and multinomial distributions for categorical variables (mice R package, version 3.19.0; \citealp{vanBuuren2011}). For non-parametric imputation, the only difference was that predictive mean matching was used for continuous variables, requiring no imputation formulae to be defined \textit{a priori}. Imputation models were trained using the reference population. The only post-processing rule bounded $alcohol$ to be non-negative. We generated five imputed datasets for each  simulation. We then fit full models for each imputed dataset. For Scenario 4, the ``full'' models omitted the variable that would ultimately be excluded (i.e. \textit{BMI}). Point estimates and standard errors were combined using Rubin’s combining rules \citep{Rubin1987}. Finally, IVW meta-analysis was run to combine estimates across the studies. Due to  consistently poor performance (see Results), imputation estimates was not generated for scenarios where the covariate structure was modified and thus we would expect additional degradation in imputation model performance (e.g. Scenarios 2 (covariate shift) or 3 (coarsened covariate)).

\section{Application: Breast cancer risk prediction}
\subsection{Reference population} \label{sec:ref_pop}
The reference population, which must include all covariates in the full model, was built using data from the National Health and Nutrition Examination Survey (NHANES). Surveys from 2015-16 and 2017-18 \citep{nhanes2015,nhanes2017} were combined using prescribed procedures (\url{https://wwwn.cdc.gov/nchs/nhanes/tutorials/Weighting.aspx}). Because family history of breast cancer and personal history of benign breast disease (BBD) were missing from the NHANES surveys, data from the National Health Interview Survey (NHIS; \citealt{nhis_data}) and from the Breast Cancer Surveillance Consortium (BCSC; \citealt{bcsc_data}) were used to build logistic regression-based imputation models using race/ethnicity and age (categorized into 10-year bins to capture non-linear effects) as predictors. Since the reference population must be complete, missingness in any other covariate was imputed using default models (predictive mean matching for continuous, logistic regression for binary, and polytomous regression for categorical) based on all other covariates except for variables that were structurally missing or related to study design. Derived variables (e.g. body mass index) were ``passively imputed'' by calculating them after imputing other variables (e.g. height and weight). Post-processing rules were specified to preserve structural missingness (e.g. nulliparous women will not have an age at first birth) and maintain internal logic  (e.g. age at menopause must be greater than age at menarche). For the purpose of this data example, missing values were singly imputed using the mice R package (version 3.19.0; \citealp{vanBuuren2011}).

\putbib          
\end{bibunit}

\newpage
\section{Supplementary Tables}

\setlength{\tabcolsep}{5pt}
\small
\begin{longtable}{lcccccccc}
\caption{Cohorts in the Breast Cancer Risk Prediction Project (BCRPP) with  recruitment information, total study participants, number of invasive breast cancer cases, and types of data collected. The polygenic risk score (PRS) column indicates participation in the Confluence genetic study; we expect to receive PRS for these cohorts. For any cohort data not currently in hand, numbers were estimated from the BCRPP platform. Recruit.: Recruitment, Mamm.: Mammographic, * indicates studies that will only share summary statistics to comply with the European Union's General Data Protection Regulation. For a list of cohort names and additional information on enrollment, see \url{https://epidataplatforms.cancer.gov/bcrpp/\#home} and \citet{brantleyPerformanceGeneralpopulationBreast2026}.}
\label{table:BCRPP} \\
\toprule
Cohort & Age at & Years of & \# & \# & Survey & Hormone & Mamm. & PRS \\
 & Recruit. & Recruit. & Females & Cases & Data & Levels & Density &  \\
\midrule
\endfirsthead
\toprule
Cohort & Age at & Years of & \# & \# & Survey & Hormone & Mamm. & PRS \\
 & Recruit. & Recruit. & Females & Cases & Data & Levels & Density &  \\
\midrule
\endhead
\midrule
\multicolumn{9}{r}{\textit{Continued on next page}} \\
\bottomrule
\endfoot
\bottomrule
\endlastfoot
AHS     & 24+      & 1993--1997 &      33{,}393 &   3{,}028 & + &   &   &   \\
BCFR    & 18+      & 1996--2012 &      33{,}982 &  13{,}133 & + & + & + & + \\
BWHS    & 21--69   & 1995       &      44{,}685 &   1{,}432 & + &   &   & + \\
CARET   & 45--74   & 1985--1994 &       6{,}289 &       507 & + &   &   &   \\
CLUE-II & 2--102   & 1989       &      17{,}372 &       956 & + &   &   &   \\
CPS-3   & 30--65   & 2006--2013 &     232{,}811 &   6{,}461 & + &   &   & + \\
CPS-II  & 40--92   & 1982       &      97{,}781 &   7{,}871 & + &   &   & + \\
CSDLH   & 21+      & 1992--1998 &       5{,}502 &   1{,}085 & + &   &   &   \\
CTS     & 22--104  & 1995--1996 &     114{,}780 &   9{,}990 & + &   &   & + \\
EPIC*   & 25--70   & 1992--1998 &     367{,}898 &  16{,}678 & + & + &   & + \\
GS*     & 16+      & 2004--2011 &     102{,}567 &  10{,}173 & + & + & + & + \\
GSSFHS* & 18--98   & 2006--2011 &      14{,}150 &       523 & + &   &   &   \\
Janus*  & 18--68   & 1974--2016 &     145{,}927 &  11{,}495 & + &   &   &   \\
MAC     & 35--75   & 2001--2014 &      19{,}787 &       549 & + &   &   &   \\
MCCS    & 40--69   & 1990--1994 &      17{,}138 &   1{,}724 & + & + &   & + \\
MEC     & 45--75   & 1993--1996 &     105{,}601 &   7{,}695 & + &   &   & + \\
MMHS    & 35--90   & 2003--2006 &       3{,}510 &   1{,}364 & + &   & + & + \\
NHS     & 30--55   & 1976       &     121{,}529 &  13{,}902 & + & + & + & + \\
NHSII   & 25--42   & 1989       &     116{,}412 &   7{,}040 & + & + & + & + \\
NYUWHS  & 34--65   & 1985--1991 &      14{,}273 &   2{,}433 & + & + &   &   \\
PLCO    & 55--74   & 1993--2001 &      76{,}110 &   5{,}998 & + & + &   & + \\
SCCS    & 40--79   & 2002--2009 &      48{,}984 &   1{,}532 & + & + &   & + \\
SIS     & 35--74   & 2004--2009 &      50{,}794 &   4{,}628 & + &   &   & + \\
SNMC*   & 18+      & 1997       &      28{,}155 &   2{,}069 & + &   &   &   \\
USRT    & 18+      & 1983--1989 &      70{,}950 &   4{,}131 & + & + &   & + \\
VITAL   & 50--76   & 2000--2002 &      40{,}340 &   1{,}870 & + &   &   &   \\
WHI     & 50--79   & 1993--1998 &     149{,}295 &  12{,}857 & + & + & + & + \\
WHS     & 45--100+ & 1993--1995 &      39{,}657 &   3{,}679 & + &   &   & + \\
WLH*    & 30--49   & 1991--1992 &      49{,}258 &   2{,}160 & + &   &   &   \\
Total   &          &            & 2{,}168{,}930 & 156{,}963 &   &   &   &   \\
\end{longtable}
\normalsize

\begin{table}[h!]
\centering
\begin{tabular}{||c c c c||} 
 \hline
 Variable name & Variable type & Distribution & Correlated with: \\ [0.5ex] 
 \hline\hline
 Age & continuous & Normal & BMI, Family history, Parity \\ 
 \hline
 BMI & continuous & Skew normal & Age, Sex, Early BMI, Parity \\
 \hline
 Early BMI & continuous & Skew normal & BMI \\
 \hline
 Height & continuous & Bimodal & Sex \\
 \hline
 Alcohol & continuous & Exponential & Sex, Smoking  \\
 \hline
 Sex & binary & Bernoulli & Smoking, BMI, Alcohol, Height \\
 \hline
 Education & categorical & Multinomial & Smoking, Parity \\
 \hline
 Family history & binary & Bernoulli & Age \\
 \hline
 Parity & discrete & Poisson & Age, Education, BMI \\
 \hline
 Smoking status & categorical & Multinomial & Sex, Education, Alcohol \\ [1ex] 
 \hline
\end{tabular}
\caption{Ten variables (five continuous, five discrete) were simulated with complex covariance patterns to produce a realistic data example. BMI = body mass index.}
\label{table:sup_sim_data}
\end{table}

\begin{table}[htbp]
\centering
\small
\begin{tabular}{lcccc}
\hline
Covariate & $\boldsymbol{\beta_1}$ & $\boldsymbol{\beta_2}$ & $\boldsymbol{\textit{OR}_1}$ & $\boldsymbol{\textit{OR}_2}$ \\
\hline
Intercept & 0 & 0 & $8.32e^{-7}$	 & $1.86e^{-7}$ \\
Age & 0.06 & 0.08 & 1.03 & 1.05 \\
BMI & 0.10 & 0.10 & 1.05 & 1.05\\
Early BMI & 0.15 & 0.15 & 1.08 & 1.08 \\
Height & 0.02 & 0.02 & 1.04 & 1.04 \\
Alcohol consumption & 0.10 & 0.20 & 1.1 & 1.2 \\
Sex & 0.40 & 0.60 & 1.2 & 1.5 \\
Education &  &  \\
\quad High school or less & Ref. & Ref. & Ref. & Ref. \\
\quad Some college & $-0.10$ & $-0.10$ & $0.9$ & $0.9$ \\
\quad College graduate or higher & $-0.20$ & $-0.20$ & $0.7$ & $0.7$\\
Family history & $1.00$ & $1.50$ & $1.5$ & $2.0$\\
Parity &  &  \\
\quad 0 & Ref. & Ref. & Ref. & Ref. \\
\quad 1 & 0.30 & 0.30 & 1.2 & 1.2 \\
\quad 2 & 0.40 & 0.40 & 1.3 & 1.3 \\
\quad 3+ & 0.40 & 0.40 & 1.3 & 1.3 \\
Smoking status &  &  \\
\quad Current & Ref. & Ref. & Ref. & Ref. \\
\quad Former & $-0.50$ & $-0.50$ & 0.8 & 0.8 \\
\quad Never & $-1.00$ & $-1.00$ & 0.5 & 0.5 \\
\hline
\end{tabular}
\caption{Regression parameters for simulations under two model specifications with smaller and larger magnitudes, respectively, for linear ($\boldsymbol{\beta_1}$, $\boldsymbol{\beta_2}$) and logistic ($\boldsymbol{\textit{OR}_1}$, $\boldsymbol{\textit{OR}_2}$) data generating processes.}. 
\label{table:sup_sim_coeffs}
\end{table}

\begin{table}[htbp]
\centering
\small
\begin{tabular}{ccccccccccccc}
\hline
Study & N & $y$ & \textit{Age} & \textit{BMI} & \textit{early BMI} & \textit{Hgt.} & \textit{Alc.} & \textit{Sex} & \textit{Edu.} & \textit{FH} & \textit{Parity} & \textit{Smok.} \\
\hline
Ref. & varied  &  & \checkmark & \checkmark & \checkmark & \checkmark & \checkmark & \checkmark & \checkmark & \checkmark & \checkmark & \checkmark \\
\hline
1           & 10000 & \checkmark & \checkmark & \checkmark & \checkmark & \checkmark & \checkmark & \checkmark & \checkmark & \checkmark &   & \checkmark  \\
2           & 20000 & \checkmark & \checkmark & \checkmark &   & \checkmark &   & \checkmark & \checkmark  &  & \checkmark & \checkmark   \\
3           & 30000 & \checkmark & \checkmark & \checkmark & \checkmark &   &   & \checkmark &   &   & \checkmark &  \checkmark  \\
4           & 8000  & \checkmark & \checkmark &   &   & \checkmark & \checkmark & \checkmark &   & \checkmark & \checkmark &    \\
5           & 50000 & \checkmark & \checkmark &   & \checkmark & \checkmark &   & \checkmark & \checkmark  &   &   &   \\
\hline
\end{tabular}
\caption{Sample sizes and covariate missingness patterns for the simulated reference population and studies. A checkmark indicates that a variable is observed, while a blank indicates it is missing by design. For sensitivity analyses, reference populations were simulated with sizes of 1,000 to 100,000. N = sample size; $y$ = outcome variable; BMI = Body mass index; Hgt. = Height; Alc. = Alcohol; Edu. = Education; FH = Family history; Smok. = Smoking status; Ref. = Reference population.}
\label{table:sup_sim_missingness}
\end{table}

\begin{table}[htbp]
\centering
\small
\begin{tabular}{cccccc}
\hline
Study & \textit{Age} & \textit{Sex} & \textit{Smok.} & corr(\textit{BMI}, \textit{early BMI}) & corr(\textit{smok.}, \textit{parity}) \\
 & (mean (var)) & (\% male) & (\%)  &  & \\
\hline
Ref. & 50 (80) & 48 & 10, 30, 60 & 0.1 & 0.2 \\
1  & 45 (10) & 55 & 15, 20, 65 & 0.3 & 0.2 \\
2  & 42 (5)  & 50 & 20, 30, 50 & 0.5 & 0.3 \\
3  & 52 (5)  & 55 & 20, 50, 30 & 0.1 & 0.1 \\
4  & 55 (10) & 45 & 10, 40, 50 & 0 & 0.1 \\
5  & 62 (5)  & 40 & 10, 50, 40 & 0.3 & 0.3 \\
\hline
\end{tabular}
\caption{Summary of \textit{age}, \textit{sex}, and \textit{smoking status} (\textit{smok.}) distributions and correlations between body mass index (\textit{BMI}) and \textit{early BMI} and between \textit{smoking status} and \textit{parity}, which were varied for the reference population and each study in Scenario 2 (covariate shift). For \textit{smoking status}, the numbers indicate the percentage of current, former, and never smokers. Ref. = Reference population.}
\label{table:sup_sim_diff_distrib}
\end{table}

\newpage
\setlength{\tabcolsep}{5pt}
\small
\begin{landscape}
\begin{longtable}{llllcccc} 
 \caption{Results from all simulation scenarios. Mean absolute percent bias (range), mean percent standard deviation (SD) (range), and mean mean-squared-error (MSE) were calculated by averaging across 100 simulations and then across covariates. Results are reported for three estimators: gold standard, the na\"ive inverse-variance weighted (IVW), and GENMETA-Lite. Options used are noted as needed. DGP = Data Generating Process; Lin. = Linear regression; Log. = Logistic regression; Covar. = Covariate; weight. = weighted reference population; partition. = partitioned covariance matrix; Select. = Selection.}
\label{table:supp_scenario_results}
\\ \hline
Scenario  & DGP & Estimator & Options & Mean abs. bias  & Mean SD  & Mean  & Fig. \\
  &  &  &  & (\%) (range) & (\%) (range) & MSE &  \\
\hline
1. Basecase & Lin. & Gold standard &  & 0.1 (-0.2, 0.3) & 2.2 (0.3,11.0) & 5.27E-05 & 3A \\
1. Basecase & Lin. & Na\"ive IVW &  & 16.4 (-8.4, 45.9) & 4.2 (0.6,19.6) & 1.67E-02 & 3A \\
1. Basecase & Lin. & Imputation & Parametric & 32.4 (-81.0, 6.8) & 3.0 (0.3,15.4) & 1.19E-01 & 3A \\
1. Basecase & Lin. & Imputation & Non-parametric & 34.6 (-85.3, 2.0) & 3.0 (0.4, 15.8) & 1.32E-01 & S5A \\
1. Basecase & Lin. & GENMETA-Lite & No weight., no partition. & 1.1 (-10.4, 0.6) & 9.8 (1.1, 46.9) & 1.15E-03 & not shown \\
1. Basecase & Lin. & GENMETA-Lite & No weight., partition. & 1.5 (-11.8, 0.5) & 10.4 (1.0, 49.4) & 1.29E-03 & 3A \\
1. Basecase & Log. & Gold standard &  & 0.7 (-1.6, 3.8) & 7.7 (0.9, 28.5) & 4.22E-04 & 3B \\
1. Basecase & Log. & Na\"ive IVW &  & 16.4 (-34.1, 74.7) & 9.5 (2.2, 31.8) & 1.52E-02 & 3B \\
1. Basecase & Log. & Imputation & Parametric & 41.6 (-83.2, -10.9) & 5.7 (0.6, 23.0) & 3.35E-02 & 3B \\
1. Basecase & Log. & Imputation & Non-parametric & 44.3 (-87.1, -10.0) & 5.6 (0.5, 21.7) & 3.68E-02 & S5B \\
1. Basecase & Log. & GENMETA-Lite & No weight., no partition. & 16.4 (-31.9, -5.3) & 14.3 (2.2, 49.9) & 4.19E-03 & not shown \\
1. Basecase & Log. & GENMETA-Lite & No weight., partition. & 16.1 (-32.5, -5.5) & 14.8 (2.5, 50.5) & 4.38E-03 & 3B \\
2. Covar. shift & Lin. & Gold standard &  & 0.1 (-0.3, 0.3) & 2.1 (0.3, 10.0) & 4.99E-05 & 4A \\
2. Covar. shift & Lin. & Na\"ive IVW &  & 20.5 (-8.8, 46.5) & 3.8 (0.5, 19.1) & 1.64E-02 & 4A \\
2. Covar. shift & Lin. & GENMETA-Lite & No weight., no partition. & 5.8 (-11.9, 12.0) & 9.4 (0.8, 55.9) & 1.95E-03 & 4A \\
2. Covar. shift & Lin. & GENMETA-Lite & Weight., no partition. & 5.6 (-14.3, 12.3) & 16.5 (1.0, 81.4) & 4.20E-03 & 4B \\
2. Covar. shift & Lin. & GENMETA-Lite & Weight., partition. & 3.2 (-6.7, 12.1) & 16.8 (1.0, 83.1) & 3.69E-03 & 4C \\
2. Covar. shift & Log. & Gold standard &  & 0.5 (-2.9, 0.4) & 7.3 (0.8, 25.8) & 3.68E-04 & 4D \\
2. Covar. shift & Log. & Na\"ive IVW &  & 21.6 (-33.2, 85.5) & 9.1 (1.9, 29.3) & 1.53E-02 & 4D \\
2. Covar. shift & Log. & GENMETA-Lite & No weight., no partition. & 15.6 (-29.0, -0.3) & 13.4 (2.4, 52.4) & 5.44E-03 & 4D \\
2. Covar. shift & Log. & GENMETA-Lite & Weight., no partition. & 16.3 (-24.7, 0.06) & 20.3 (2.6, 74.3) & 7.03E-03 & 4E \\
2. Covar. shift & Log. & GENMETA-Lite & Weight., partition. & 13.6 (-25.1, -0.2) & 20.7 (2.6, 75.4) & 6.14E-03 & 4F \\
3. Coarse covar. & Lin. & Gold standard &  & 0.2 (-1.6, 0.4) & 2.2 (0.3, 11.5) & 5.30E-05 & S6A, S6B \\
3. Coarse covar. & Lin. & Na\"ive IVW & Without coarse estimate & 18.0 (-8.6, 45.1) & 4.1 (0.6, 18.5) & 1.67E-02 & S6A \\
3. Coarse covar. & Lin. & Na\"ive IVW & With coarse estimate & 16.4 (-8.6, 45.1) & 4.1 (0.6, 18.5) & 1.67E-02 & S6B \\
3. Coarse covar. & Lin. & GENMETA-Lite & No weight., partition. & 1.2 (-6.0, 0.9) & 10.5 (1.1, 55.1) & 1.25E-03 & S6A, S6B \\
3. Coarse covar. & Log. & Gold standard &  & 0.7 (-0.4, 3.3) & 8.1 (1.0, 30.6) & 4.81E-04 & S6C, S6D \\
3. Coarse covar. & Log. & Na\"ive IVW & Without coarse estimate & 16.1 (-34.1, 80.7) & 9.4 (1.9, 31.8) & 1.60E-02 & S6C \\
3. Coarse covar. & Log. & Na\"ive IVW & With coarse estimate & 16.8 (-34.1, 80.7) & 9.3 (1.9, 31.8) & 1.60E-02 & S6D \\
3. Coarse covar. & Log. & GENMETA-Lite & No weight., partition. & 16.0 (-28.8, -5.7) & 14.7 (2.5, 57.0) & 4.10E-03 & S6C, S6D \\
4. Model select. & Lin. & Gold standard &  & 0.3 (-1.4, 0.4) & 2.3 (0.3, 11.1) & 6.01E-05 & 5A \\
4. Model select. & Lin. & Na\"ive IVW &  & 16.8 (-14.4, 48.7) & 4.7 (0.6, 22.0) & 1.82E-02 & 5A \\
4. Model select. & Lin. & Imputation & Parametric & 33.1 (-81.4, -3.5) & 3.1 (0.3, 16.4) & 1.29E-01 & 5A \\
4. Model select. & Lin. & Imputation & Non-parametric & 35.9 (-85.2, -4.8) & 3.3 (0.4, 16.9) & 1.42E-01 & not shown \\
4. Model select. & Lin. & GENMETA-Lite & No weight., partition. & 1.1 (-8.3, 0.4) & 10.8 (1.0, 49.3) & 1.38E-03 & 5A \\
4. Model select. & Log. & Gold standard &  & 1.1 (-3.2, 1.4) & 12.7 (1.1, 46.2) & 1.16E-03 & 5B \\
4. Model select. & Log. & Na\"ive IVW &  & 22.8 (-41.9, 102.3) & 15.8 (2.9, 55.0) & 3.10E-02 & 5B \\
4. Model select. & Log. & Imputation & Parametric & 41.2 (-82.5, 8.7) & 9.7 (1.3, 39.7) & 3.44E-02 & 5B \\
4. Model select. & Log. & Imputation & Non-parametric & 44.3 (-86.0, 14.3) & 9.9 (0.9, 38.2) & 3.79E-02 & not shown \\
4. Model select. & Log. & GENMETA-Lite & No weight., partition. & 13.2 (-35.7, 2.3) & 19.8 (3.0, 66.1) & 4.78E-03 & 5B \\

\bottomrule{}
 \end{longtable}
 \normalsize
 \end{landscape}

\newpage
\section{Supplementary Figures}
 
\begin{figure}[!htb]
    \centering
    \includegraphics[height=1\textwidth]{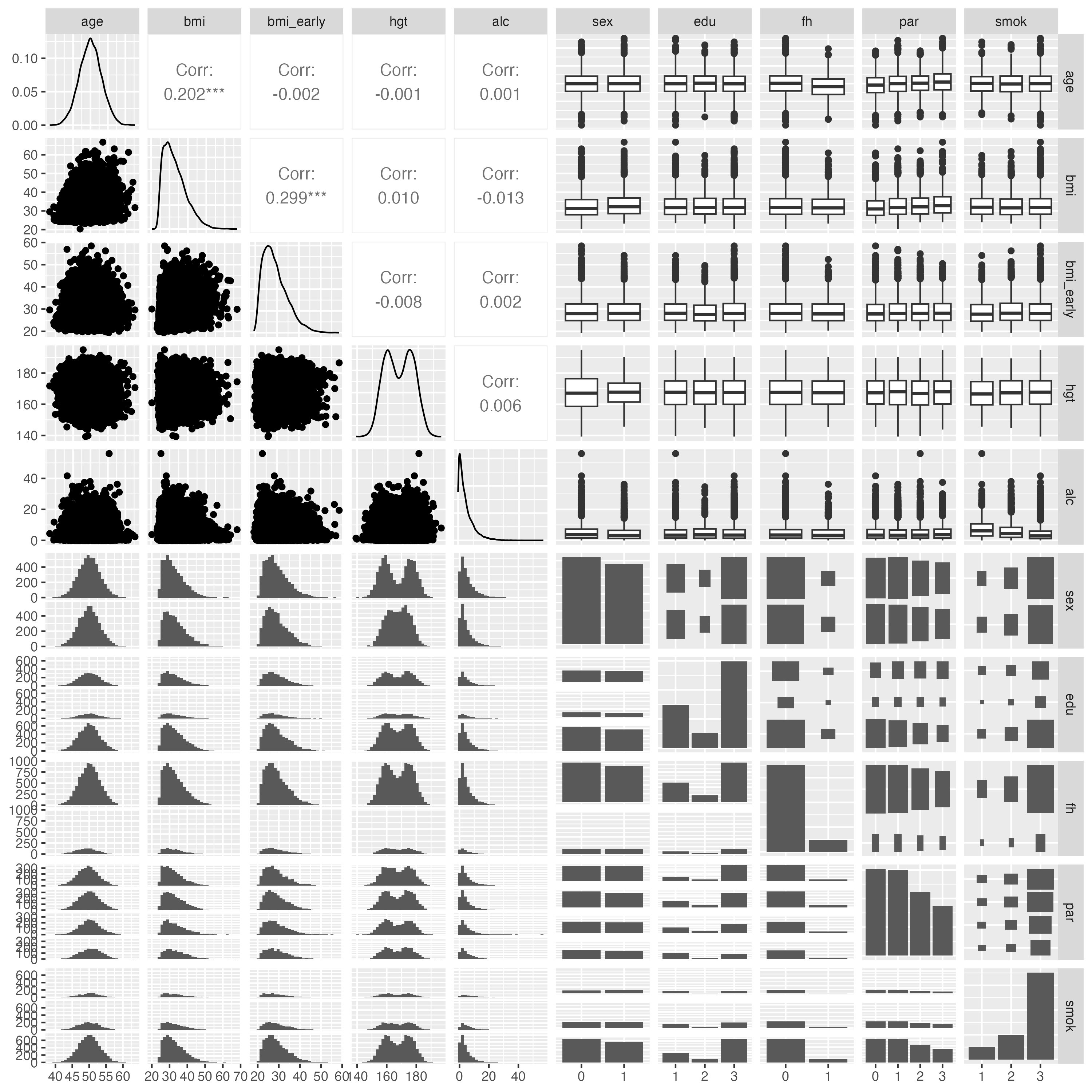}
    \caption{Distributions of and correlations among simulated covariates. BMI = body mass index; Hgt. = Height; Alc. = Alcohol; Edu. = Education; FH = Family history; Par. = Parity; Smok. = Smoking status; Corr. = Pearson's correlation coefficient.}
    \label{fig:sup_sim_ggpairs}
\end{figure}

\begin{figure}[!htb]
    \centering
    \includegraphics[height=0.6\textwidth]{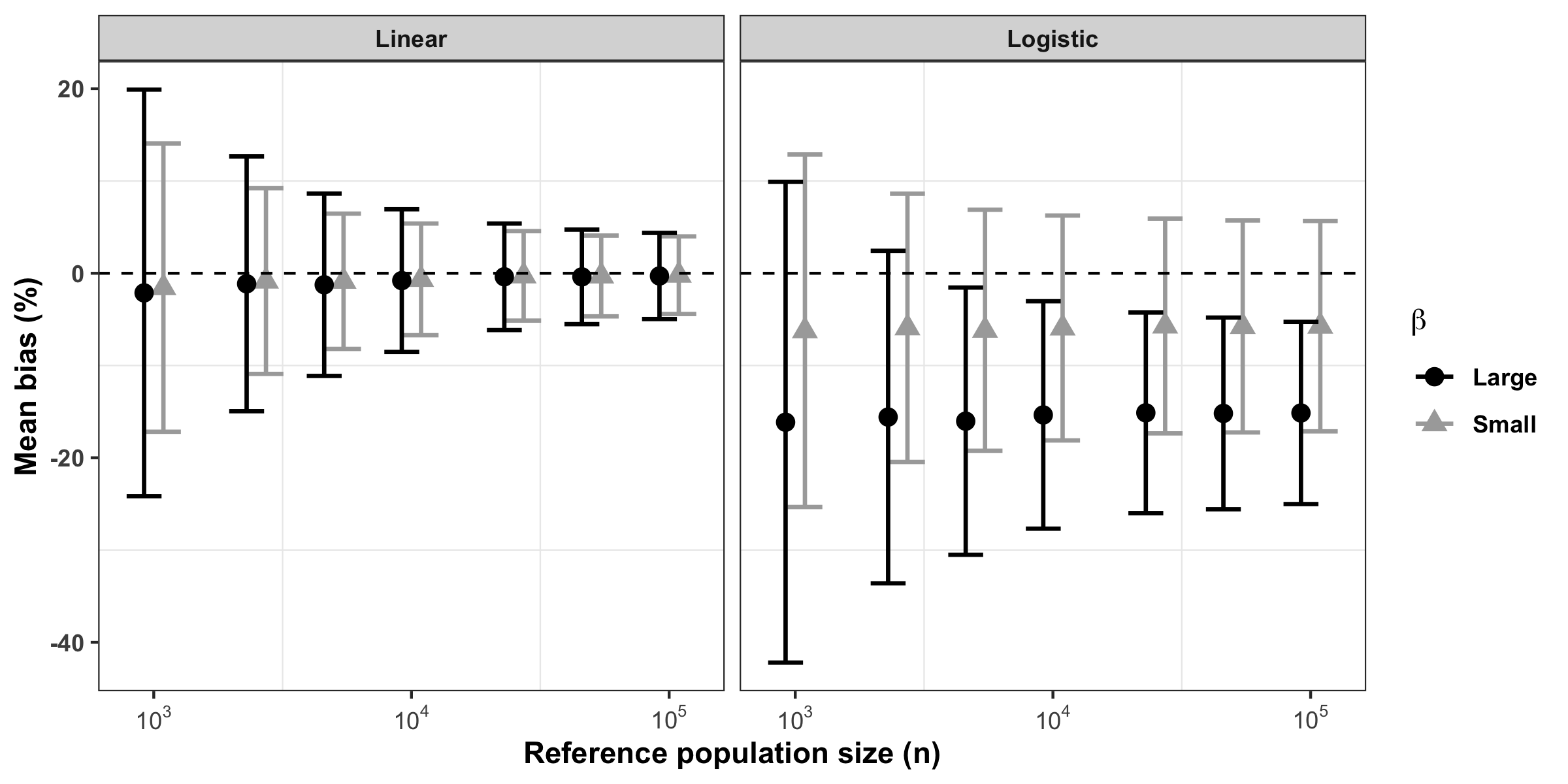}
    \caption{Effect on the GENMETA-Lite estimator's bias and variance of altering the sample size of the reference population ($n_{\text{ref}}=$ 1,000 to 100,000) and the magnitude of the regression coefficients (larger $\boldsymbol{\beta}$ in black circles, smaller $\boldsymbol{\beta}$ in grey triangles; see Table \ref{table:sup_sim_coeffs}), for (A) linear and (B) logistic data generating processes. Mean percent bias and empirical standard deviation (represented by error bars) were calculated by averaging percent bias and standard deviation for each coefficient estimate across 100 simulations, and then averaging over all covariates.}
    \label{fig:sens_analy_bias_sd}
\end{figure}

\begin{figure}[!htb]
    \centering
    \includegraphics[height=0.9\textwidth]{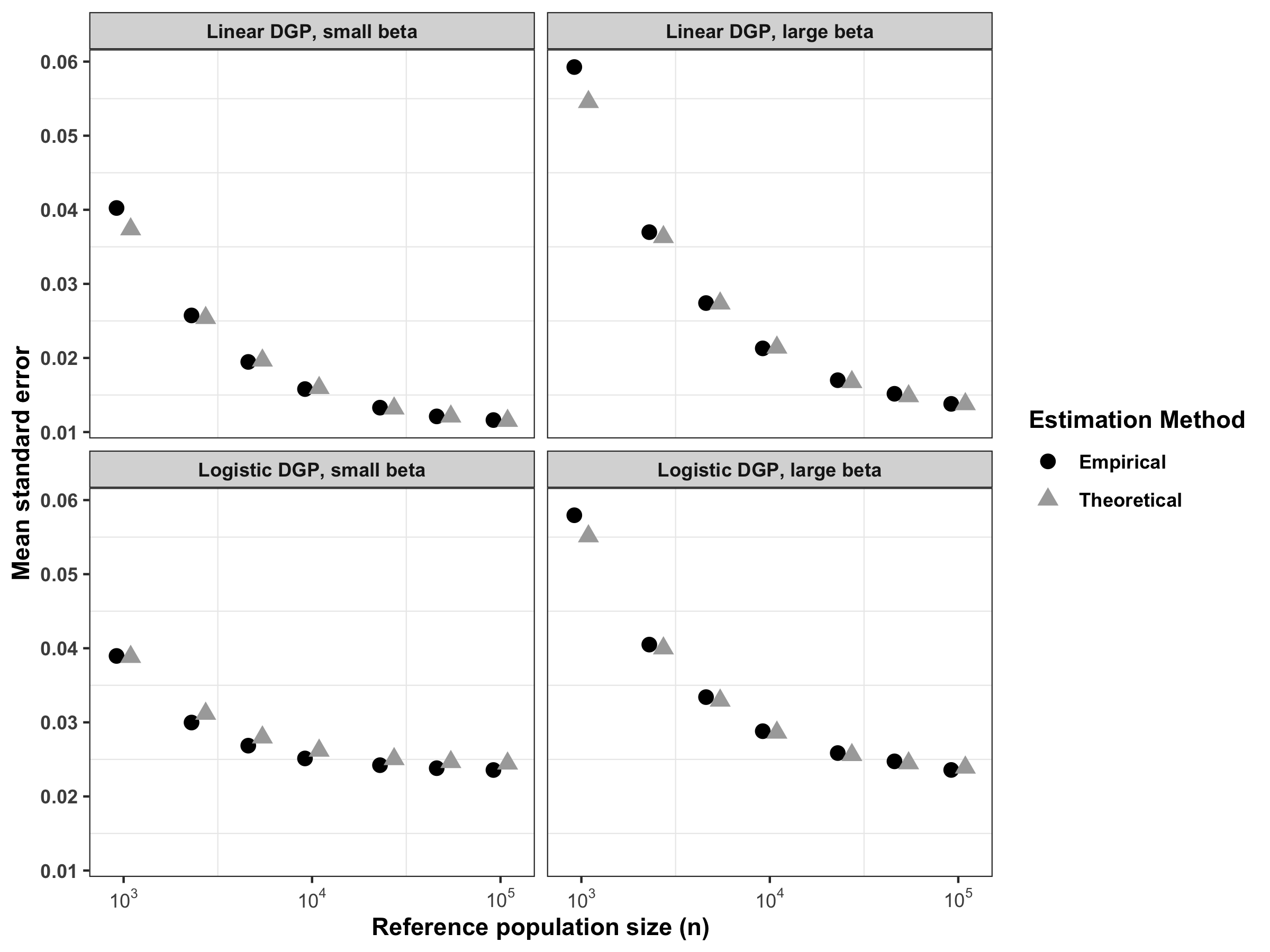}
    \caption{Effect on the GENMETA-Lite estimator's mean empirical standard deviation (SD; black circles) and theoretical standard error (SE; grey triangles) of varying the sample size of the reference population ($n_{\text{ref}}=$ 1,000 to 100,000), the magnitude of the regression coefficients (smaller $\boldsymbol{\beta}$ in left column; larger $\boldsymbol{\beta}$ in right column; see Table \ref{table:sup_sim_coeffs}), and the data generating process (DGP; linear on top row; logistic on bottom row). Mean empirical SD was obtained by computing the SD for each covariate across 100 simulations and averaging over all covariates. Mean theoretical SE was calculated averaging SEs for each coefficient  across 100 simulations, and then averaging over all covariates.} \label{fig:sens_analy_theor_se_emp_sd}
\end{figure}

\begin{figure}[!htb]
    \centering
    \includegraphics[height=0.6\textwidth]{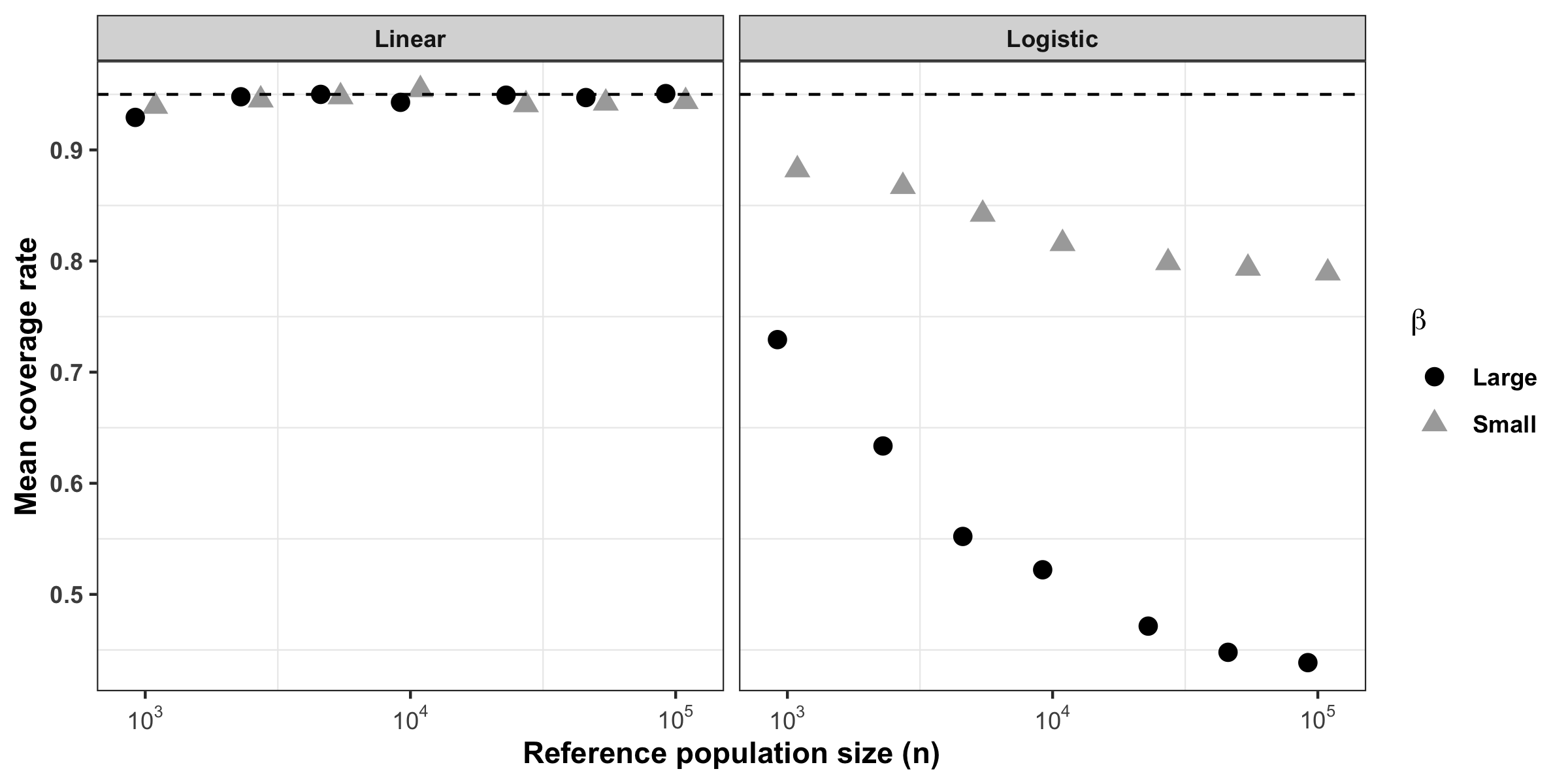}
    \caption{Effect on the GENMETA-Lite estimator's 95\% coverage rates of altering the sample size of the reference population ($n_{\text{ref}}=$ 1,000 to 100,000) and the magnitude of the regression coefficients (larger $\boldsymbol{\beta}$ in black circles, smaller $\boldsymbol{\beta}$ in grey triangles; see Table \ref{table:sup_sim_coeffs}), for (A) linear and (B) logistic data generating processes. Mean coverage rates were calculated by averaging coverage rates for each coefficient estimate across 100 simulations, and then averaging over all covariates.}
    \label{fig:sens_analy_coverage_rates}
\end{figure}

\begin{figure}[!htb]
    \centering
    \includegraphics[height=0.6\textwidth]{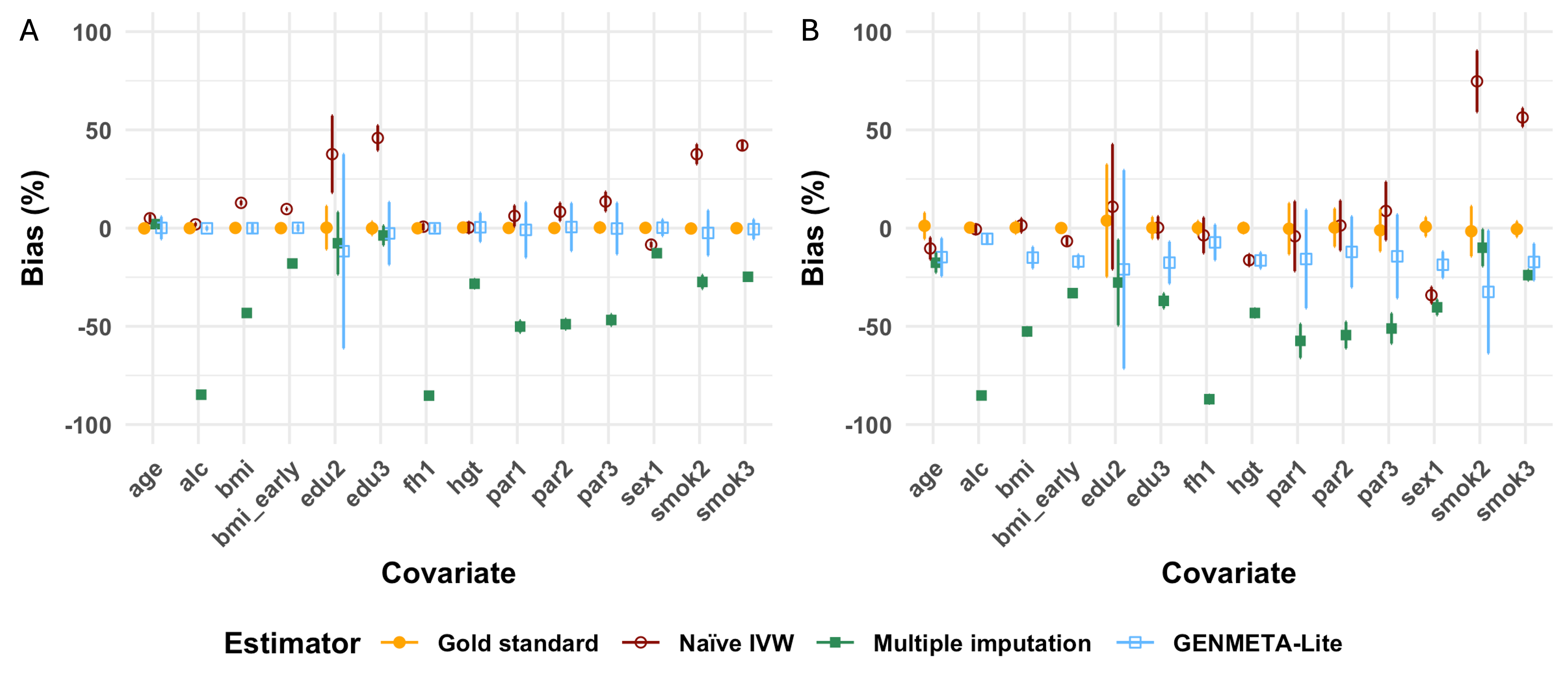}
    \caption{Results for Scenario 1 (basecase) for mean percent bias under (A) a linear data generating process (DGP) and (B) a logistic DGP, for the gold standard estimator (filled gold circles), the na\"ive inverse-variance weighted (IVW) estimator (open red circles), multiple imputation estimator with non-parametric imputation (M=5; filled green squares), and GENMETA-Lite estimator (open blue squares); error bars indicate empirical standard deviation of percent bias across simulations. Larger regression coefficients were used to simulate the outcome variables (see $\boldsymbol{\beta_2}$ and $\boldsymbol{\textit{OR}_2}$ in Table \ref{table:sup_sim_coeffs}). A partitioned covariance matrix, which combined covariance estimates from each study and covariance estimates from an unweighted reference population ($n_{\text{ref}} = 5000$), was used to estimate covariance across all covariates.}
    \label{fig:basecase_logistic_pmm_partitioned_var}
\end{figure}

\begin{figure}[!htb]
    \centering
    \includegraphics[height=1\textwidth]{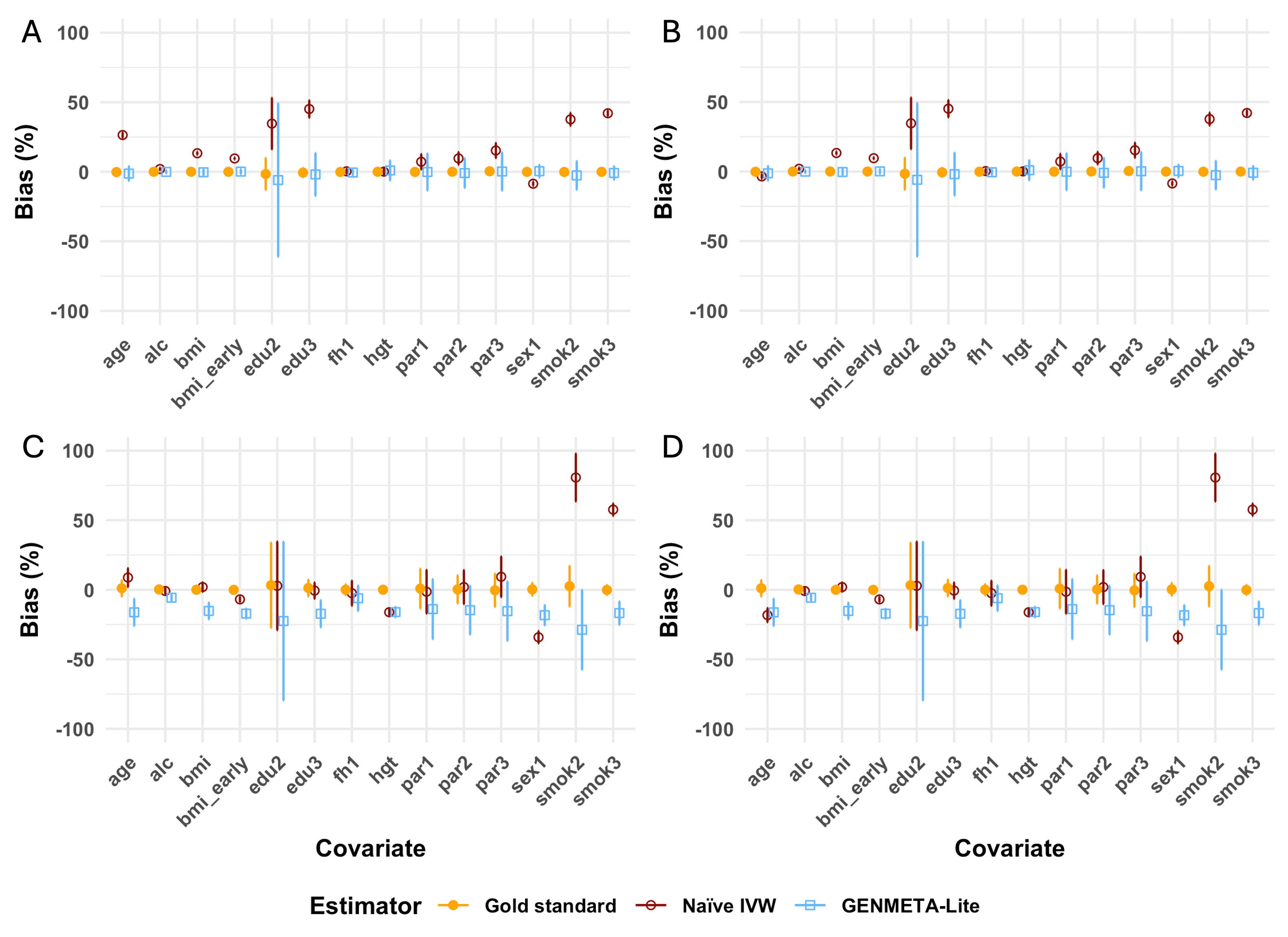}
    \caption{Results for Scenario 3 where \textit{age} was coarsened to 5-year bins in two out of five studies. Mean percent bias under linear (A, B) and logistic (C, D) data generating processes for the gold standard estimator (filled gold circles), the naive inverse-variance weighted (IVW)  estimator (open red circles), and the GENMETA-Lite estimator (open blue squares); error bars indicate empirical standard deviation of percent bias across simulations. IVW estimates for the coarsened covariate \textit{age} differed slightly when the coarsened covariate estimate was excluded (A, C) versus included (B, D) with the estimates for the continuous \textit{age} covariates. Larger regression coefficients were used to simulate the outcome variable (see $\boldsymbol{\beta_2}$ and $\boldsymbol{\textit{OR}_2}$ in Table \ref{table:sup_sim_coeffs}). A partitioned covariance matrix, which combined covariance estimates from each study and covariance estimates from an unweighted reference population ($n_{\text{ref}} = 5000$), was used to estimate covariance across all covariates.}
    \label{fig:coarsened}
\end{figure}

\begin{figure}[!htb]
    \centering
    \includegraphics[height=1\textwidth]{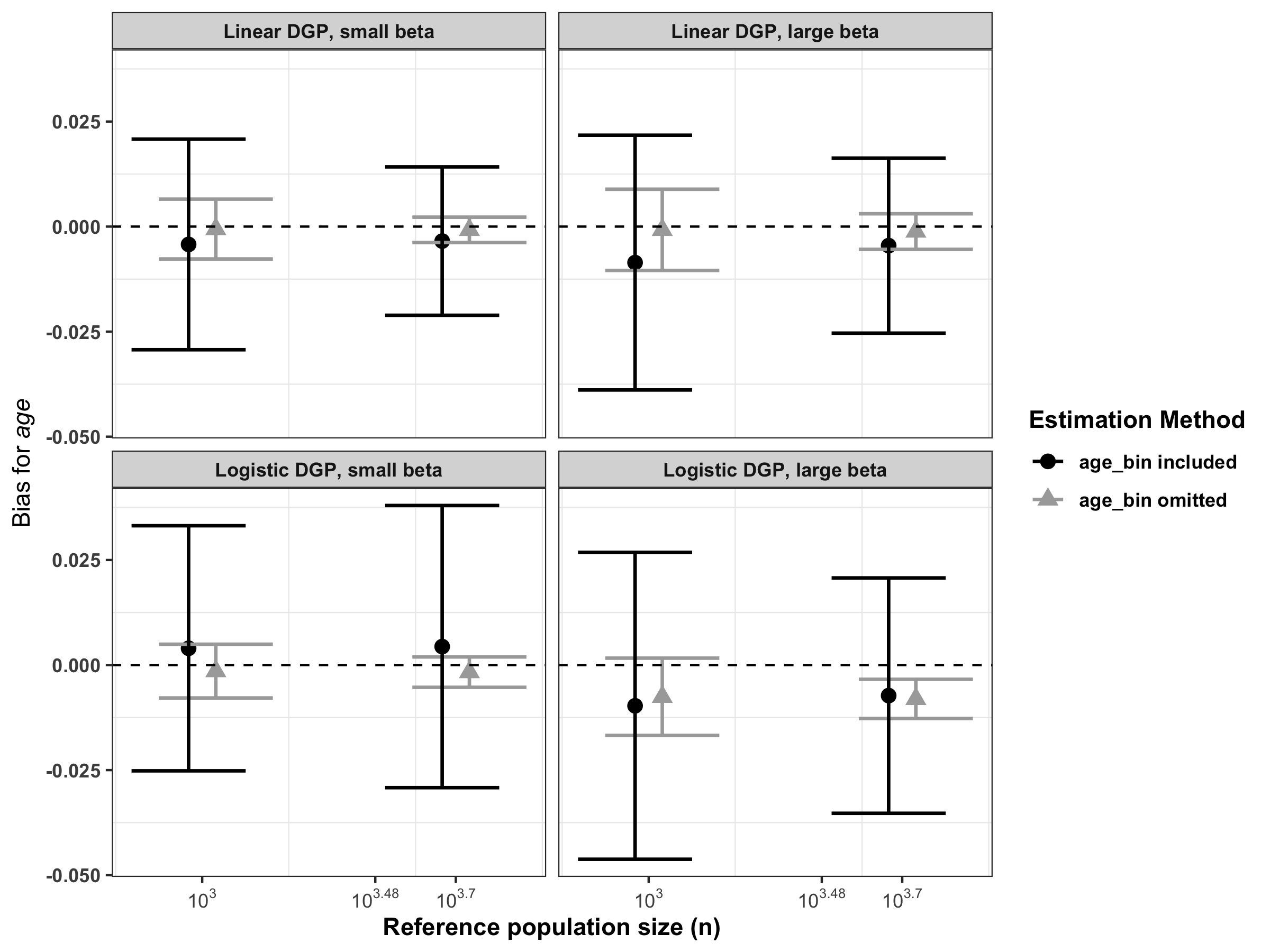}
    \caption{Sensitivity analyses for Scenario 3, where \textit{age} was coarsened to 5-year bins in two out of five studies, exploring the effect on the GENMETA-Lite estimator's bias and empirical standard deviation (represented by error bars) of the coefficient estimate for $age$ when the coarsened variable (\textit{age\_bin}) was included versus excluded from the final model for linear (top row) and logistic (bottom row) data generating processes using two different reference population sizes ($n_{\text{ref}} = 1000$ and $n_{\text{ref}}=5000$), and two coefficient magnitudes (see Table~\ref{table:sup_sim_coeffs}).}
    \label{fig:coarsened_incl_vs_omitted_bias_sd}
\end{figure}

\begin{figure}[!htb]
    \centering
    \includegraphics[height=1\textwidth]{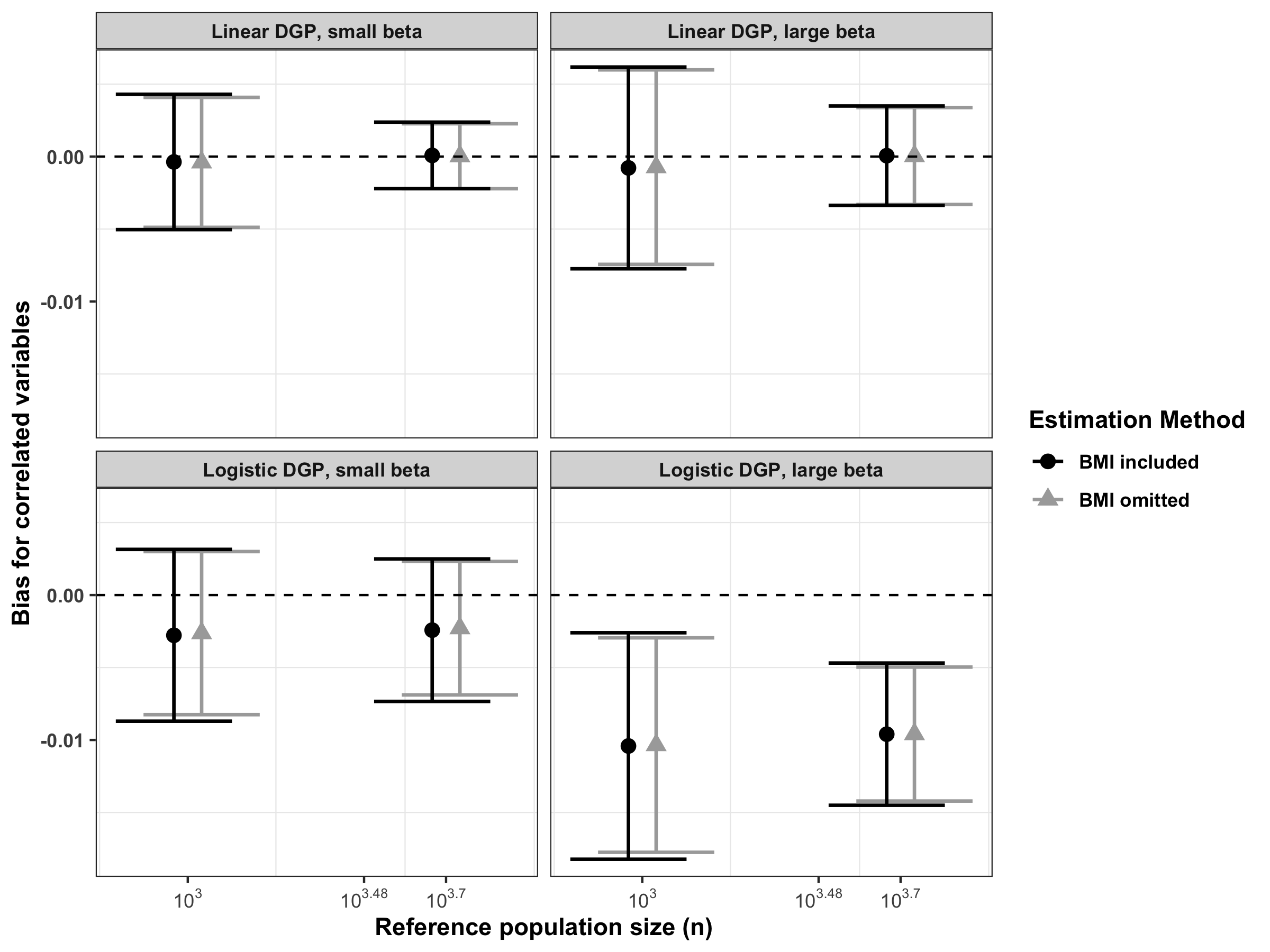}
    \caption{Sensitivity analyses for Scenario 4, where \textit{BMI} not included in the data generating process but was fit in study-specific models, showing the effect on the GENMETA-Lite estimator's bias and empirical standard deviation (represented by error bars) of the coefficient estimates for variables strongly correlated variables with \textit{BMI} (i.e., \textit{age} and \textit{early BMI}) when \textit{BMI} was included versus excluded from the final model under linear (top row) and logistic (bottom row) data generating processes (DGP) using two different reference population sizes ($n_{\text{ref}} = 1000$ and $n_{\text{ref}}=5000$), and two coefficient magnitudes (see Table~\ref{table:sup_sim_coeffs}).}
    \label{fig:model_sel_incl_vs_omitted_bias_sd}
\end{figure}

\begin{figure}[!htb]
    \centering
    \includegraphics[height=1\textwidth]{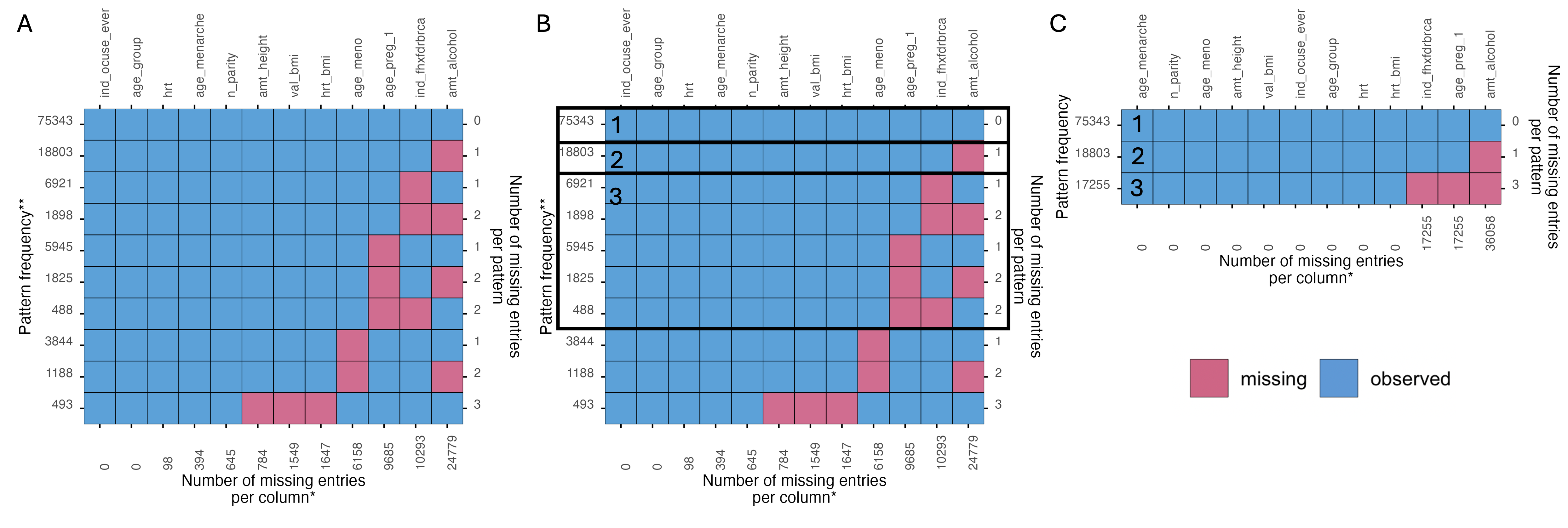}
    \caption{Example workflow to handle data that are missing at random (MAR) and missing by design within a cohort. Initially (A) all missingness patterns were plotted (ggmice R package); only the top 10 most common missingness patterns are shown here (red = missing covariates; blue = observed covariates). Then (B) common missingness patterns were combined, minimizing the number of missing covariates per group, until two to four subgroups of individuals covered \textgreater90\% of the full study population. Regression models were fit within each group (C), and coefficient estimates were combined across groups  using GENMETA-Lite to obtain final estimates for a ``full'' model including all covariates observed in that cohort.}
    \label{fig:id_dominant_missingness_patterns}
\end{figure}

\begin{figure}[!htb]
    \centering
    \includegraphics[height=1\textwidth]{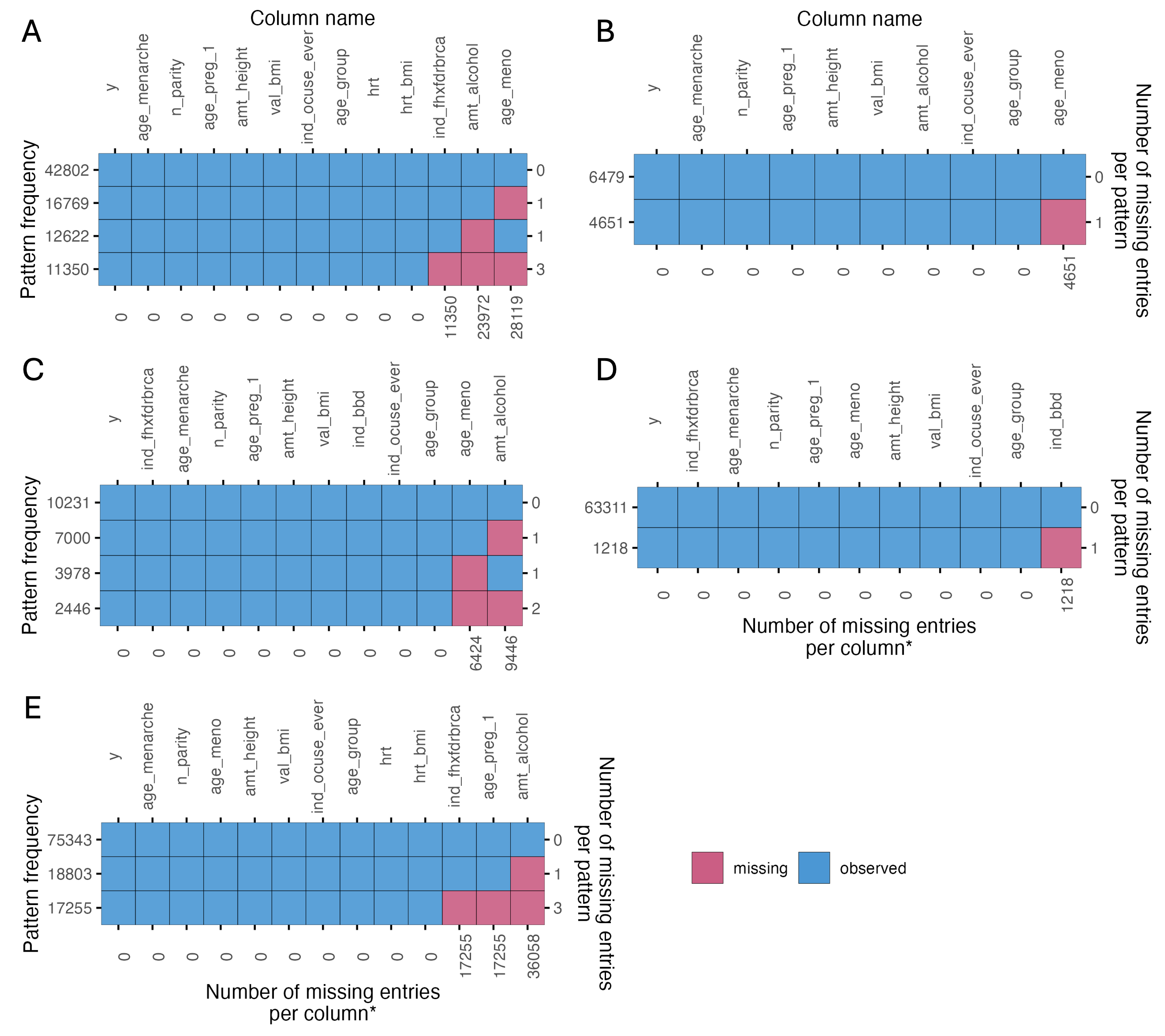}
    \caption{Dominant missingness patterns used to fit regression models in (A) Cancer Prevention Study-3 (CPS3), (B) Melbourne Collaborative Cohort Study (MCCS), (C) Nurses' Health Study (NHS), (D) Prostate, Lung, Colorectal, and Ovarian Cancer Screening Trial (PLCO), and (E) Women's Health Initiative (WHI) cohorts. Each row denotes a reduced model that was fit with the ``observed'' (blue) covariates while the ``missing'' (red) covariates were excluded from the model. }
    \label{fig:nested_model_missingness}
\end{figure}

\begin{figure}[!htb]
    \centering
    \includegraphics[height=1\textwidth]{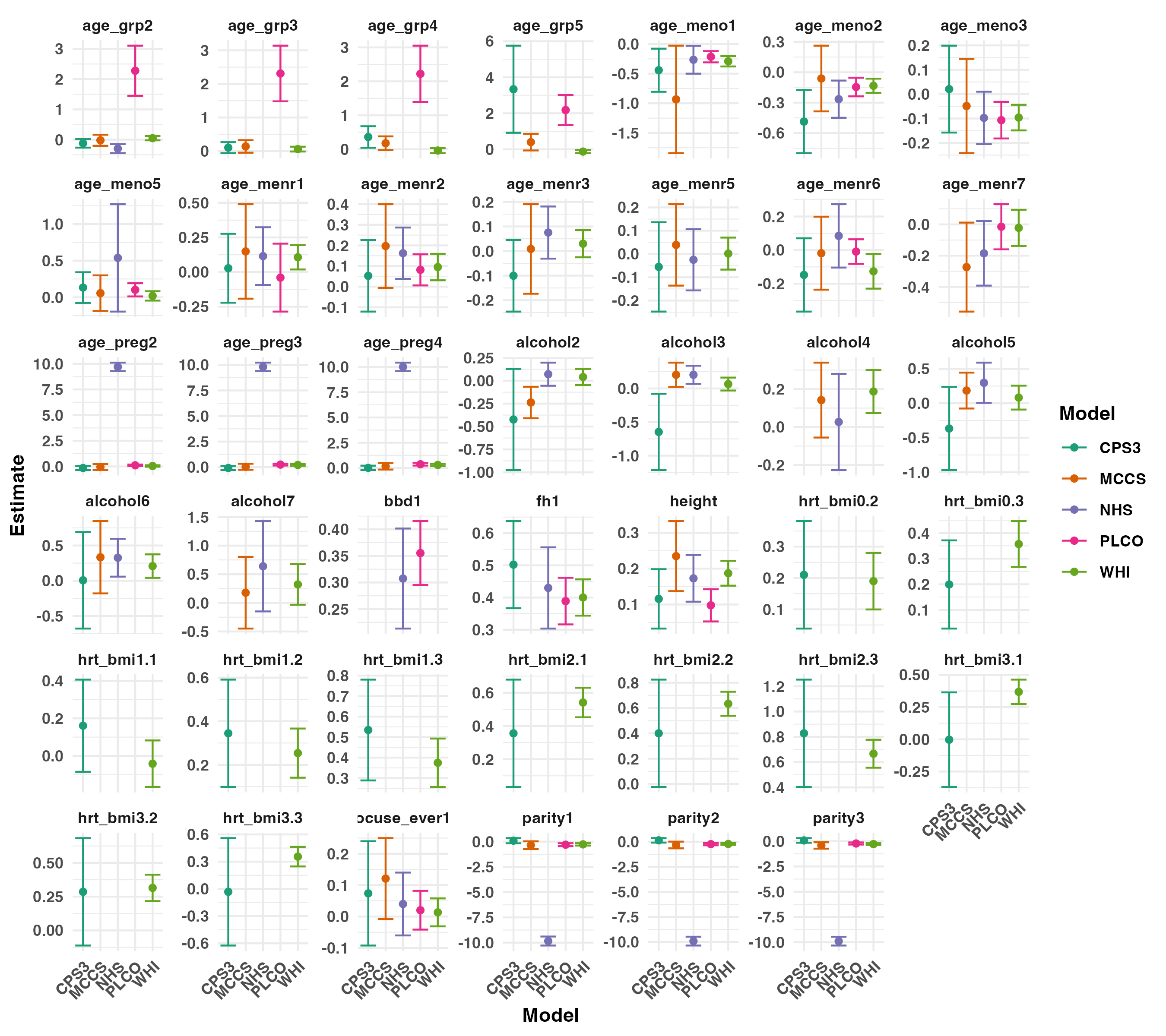}
    \caption{Coefficient estimates for each cohort obtained using GENMETA-Lite. Missing estimates indicate that the covariate was not available in that cohort. Error bars show 95\% confidence intervals. CPS3 = Cancer Prevention Study-3; MCCS = Melbourne Collaborative Cohort Study; NHS = Nurses' Health Study; PLCO = Prostate, Lung, Colorectal, and Ovarian Cancer Screening Trial; WHI = Women's Health Initiative; BBD = benign breast disease; BMI = body mass index; HRT = hormone replacement therapy; OC = oral contraceptive.}
    \label{fig:RWD_est_across_cohorts}
\end{figure}




\end{document}